\documentclass[preprint,onecolumn,showpacs,showkeys,aps,superscriptaddress,floatfix]{revtex4-1}
\usepackage{amsmath}
\usepackage{amssymb}
\usepackage{graphicx}
\usepackage{esint}
\usepackage[utf8]{inputenc}
\makeatletter

\usepackage{amssymb,amsthm}
\renewcommand{\vec}[1]{\boldsymbol{#1}}
\newcommand{\braOket}[3]{\left\langle#1\middle|#2\middle|#3\right\rangle}
\newcommand{\ket}[1]{\left|#1\right\rangle}
\newcommand{\bra}[1]{\left\langle#1\right|}

\@ifundefined{showcaptionsetup}{}{%
 \PassOptionsToPackage{caption=false}{subfig}}
\usepackage{subfig}
\makeatother

\begin{document}

\title{Electronic correlations in the Hubbard model on a bi-partite lattice}

\author{Wissam A. Ameen}

\affiliation{Theoretical Physics Division, School of Physics and Astronomy, University
of Manchester, Manchester M13 9PL, United Kingdom}

\affiliation{Physics Department, College of Science, University of Anbar, Anbar, Iraq}

\author{Niels R. Walet}
\affiliation{Theoretical Physics Division, School of Physics and Astronomy, University
of Manchester, Manchester M13 9PL, United Kingdom}

\author{Yang Xian}

\affiliation{Theoretical Physics Division, School of Physics and Astronomy, University
of Manchester, Manchester M13 9PL, United Kingdom}

\begin{abstract}
In this work we study the Hubbard model on a bi-partite lattice using
the coupled-cluster method (CCM). We first investigate how to implement
this approach in order to reproduce the lack of magnetic order  in
the 1D model, as predicted by the exact Bethe-Ansatz solution. 
This result can only be reproduced if we include an algebraic
correlation in some of the coupled-cluster model coefficients.
Using the correspondence between the Heisenberg model and the Hubbard model in the
large-coupling limit, we use  very
accurate results for the CCM applied to the Heisenberg and its generalisation, the $XXZ$ model,
to determine CCM coefficients with the correct properties.
Using the same approach we then study the 2D Hubbard model on a square and
a honeycomb lattice, both of which can be though of as simplified models of real 2D
materials. We analyse the charge and spin excitations,
 and show that with care we can obtain good results.
\end{abstract}



\maketitle

\section{Introduction}

The Hubbard model \cite{hubbard1963electron} and its variations have
been widely applied to investigate the electronic correlations of
interacting electrons in low-dimensional systems. The simple form of
the model not only provides an excellent test ground for bench-marking
theoretical tools, but also has important applications in describing
experimental data. The model has played an important role in our
understanding of the high-$T_{c}$ superconductors over the last few
decades, see, e.g., Ref.~\cite{hirsch1985attractive}, and has been extensively studied
using many microscopic methods, see, e.g., Refs.\ \cite{avella_strongly_2012,avella_strongly_2013} for recent reviews.

The Hubbard model consists of only two terms: a nearest-neighbour
electron hopping term with strength $t$ and an on-site electronic
repulsion with strength $U$. In two dimensions on a square lattice
when half the available electronic states are filled, the on-site
repulsion causes a Mott transition from a paramagnetic conductor to an
antiferromagnetic insulator for any non-zero value of $U$
\cite{hirsch1985attractive}.  However, the model on a honeycomb
lattice shows a different picture: the paramagnetic state is stable
for small $U$, and the Mott transition occurs at a non-zero value of
the interaction $U=U_{c}$, which has a value of
about $U_{c}/t\approx4.5$ (see e.g., Ref.~\cite{he2012cluster}). This
quantum phase transition has attracted strong theoretical interests
since the discovery of graphene and other two-dimensional materials
such as silicene and Boron Nitride \cite{xu2013graphenelike} due to
their hexagonal structures. Of course, the application of the Hubbard model to
graphene and sister materials may be questioned and is a subject of
ongoing debate; most theoretical studies of interaction effects in
graphene employ the full long-range Coulomb interaction
\cite{kotov2012electronelectron}. Nevertheless, the Hubbard model with
local interactions (the on-site and nearest neighbour interactions) has
been used to investigate the electronic correlations in
graphene such as the possible edge magnetism of narrow ribbons and
formation of local magnetic moments (see references in
\cite{kotov2012electronelectron}). Furthermore, there is also some
theoretical discussion of possible spin-liquid phase between the
metallic phase and the ordered antiferromagnetic insulating phase on a
honeycomb lattice. Sorella \emph{et al} \cite{sorella2012absence}
report, using a numerically exact Monte Carlo method, that there is
little or no indication of such a phase transition to a spin liquid in
clusters of 2592 atoms. Both Sorella \emph{et al} and He \emph{et al}
 \cite{he2012cluster} argue
for a spin-liquid state below $U_{c}$, with a semi-metallic state only
at $U=0$, with evidence for a first order Mott phase transition.  Yang
\emph{et al} \cite{yang2012effective} apply an effective Hamiltonian
approach, and also find weak or no evidence for a quantum spin liquid.
In the work of Lin \emph{et al }\cite{lin2015phasetransitions}, a
slightly modified version of the Hubbard model is studied.

One of the important methods to systematically study  electronic
correlations of interacting electron systems is the coupled-cluster
method (CCM) \cite{bishop1991anoverview}. The key advantages of the CCM are its
avoidance of unphysical divergences in the thermodynamic limit and its ability
to be taken to high accuracy through systematic inclusion of high-order correlations. 
The price to pay is that the method does not provide a variational bound to the
ground state energy, and that the wave function is not Hermitian.
Nevertheless, convergence of CCM calculation has been found to be rapid.
Therefore, the CCM is the method of choice for state-of-the-art calculations for atomic and molecular 
systems in quantum chemistry, where it is used, amongst others, to calculate correlation energies,
 to an accuracy of less than one millihartree (1 mH) \cite{bartlett_coupled-cluster_2007}. The CCM has been successfully applied to a wide range of other physical systems, including problems in nuclear physics, both for finite and infinite
nuclear matter, the electron gas and liquids, as well as various integrable and nonintegrable models,
and various relativistic quantum field theories. In most such cases the numerical
results are either the best or among the best available. A classical example  is the
electron gas, where the coupled cluster results for the correlation energy agree over the entire
metallic density range to within less than 1 millihartree (or less than 1\%) with the essentially exact Green's function Monte Carlo results \cite{bishop1991anoverview}. Most relevant to our present work, over the last two decades the CCM has also been successfully applied to describe quantum spin lattice systems accurately, providing some of the best numerical results for the ground state energy, the sub-lattice magnetization, and spin-wave excitation spectra (for a recent example see, e.g., Ref.~\cite{li2015groundstate}). In these applications, the full advantage of the systematic improvement by inclusion higher-order correlations attainable by computer algebra have revealed physical properties of the quantum phase transitions in spin systems.

The CCM shares quite a few of its roots with classic many-body techniques based
on  many-body perturbation theory, see
 Refs.~\cite{avella_strongly_2012,avella_strongly_2013} for some modern examples. 
The result of CCM calculations look much like a resummation of the perturbation series, and indeed do not diverge where
perturbation theory fails to converge.
It is
a pure method: normally one works directly with the full complement of quasi-particles relative to a generalised vacuum --usually called reference-- state.
There is a similarity with dynamical mean-field theory. The lowest order of CCM
is like mean-field theory, and one can include
higher order RPA-like correlations. In the standard formulation applied here
it lacks the full power of the normal mean-field,
which is included in the dynamical mean-field method, but including higher order
correlations is much more systematic in CCM. Clearly the CCM is one of the family
of cluster approximations. In its calculations all independent cluster
excitations on the reference wave function are taken into account in the 
ket state--but not in the bra state to make the calculations practical. 
There is a variational formulation of the CCM, but with an independent bra and ket state variation,  there is no variational upper bound for the 
energy, and excited states are solutions to a non-Hermitian eigenvalue problem.
On the other hand, one can show that the Helmann-Feynman theorem is satisfied.
It also means that we cn easily evaluate expectation values of any observable.
Thus using CCM has both advantages and disadvantages relative to other
many-body methods, and deserves to be studied in more detail for the Hubbard model. 

There are a few CCM calculations using very simple approximations for the Hubbard model on
a square lattice \cite{roger1990thecoupledcluster,
  bishop1995amicroscopic}. The nature of these calculations 
involves a systematic cluster truncation
  of the wave function;
  here we include higher-order
correlations than in previous work, and extend the calculations to the honeycomb
lattice. We take advantage of the fact that the Hubbard model
reduces to a spin model in the large-$U$ limit and employ the existing
CCM results for lattice spin models to obtain better numerical results
for the ground-state energy and the sub-lattice magnetization.
These ideas show some similarity with the work by Zheng, Paiva and collaborators
\cite{paiva_ground-state_2005,zheng_magnon_2005}, who study the Hubbard model for large $U$ using a series expansion technique, and make use of the
Heisenberg model results as well.

This paper is organized as follows. In Sec.~II we introduce the
Hubbard model and discuss its relation to the spin models. In Sec. III
we provide a brief description of the CCM and the detail of its
application to the Hubbard model for the ground state and excited
states, including both the charge and spin-flip excitations. A
particular high-order approximation scheme employing the earlier CCM
results is introduced and applied. We also give the consistent results
for the Hubbard model in the large-$U$ limit with the spin models. In
the results section of Sec.~\ref{sec:Results}, we summarize all the
results for the 1D chain, and the 2D square and honeycomb models. We
emphasize the significant improvement for a wide range of values of
$U$ in the numerical results for the ground state energies and the
magnetic order parameter (sub-lattice magnetisation) when the high-order correlations are included. We
include a discussion of the indication of a phase transition for the
honeycomb lattice. In the last section we provide a summery of our
results and a discussion on the technical difference between the CCM
calculations for the Hubbard and spin models. 
Some of the details of the CCM calculations can be found in appendix \ref{app:A}.
Since our work relies
on the correspondence between the Hubbard and Heisenberg models, we
discuss some pertinent details of that correspondence in appendix \ref{app:B}.

\section{The Hubbard Model\label{sec:The-Hubbard-Model}}

We start from the Hubbard model defined on a bi-partite lattice,
consisting of a hopping term with strength $t$ and an on-site
potential $V$ with strength $U$ in terms of electron-creation operators $c^\dagger_{\vec{m} \sigma}$,
\begin{eqnarray}
H&=&-t\sum_{\langle\vec{i}\vec{j}\rangle \sigma}\left(c_{\vec{i} \sigma}^{\dagger}c_{\vec{j} \sigma}+c_{\vec{j} \sigma}^{\dagger}c_{\vec{i} \sigma}\right)+
U
\sum_{\vec{m}}\left(n_{\vec{m}\uparrow}-\frac{1}{2}\right)\left(n_{\vec{m}\downarrow}-\frac{1}{2}\right)+\frac{NU}{4}
\label{eq:Hubbard}\\
&\equiv&-t T+U V,
\end{eqnarray}
where the index $\vec{m}$ runs all $N$ lattice sites. The notation
$\langle\vec{i}\vec{j}\rangle$ denotes a sum over nearest-neighbour
sites (which are by definition on opposite sub-lattices), and we shall
use indices $\vec{i}$ and $\vec{j}$ exclusively for one of the two sub-lattices,
called $A$ and $B$, respectively. The spin index $\sigma=\uparrow,\downarrow$.
Finally, the parameters $t$ and $U$ are the hopping and on-site
interaction strengths, respectively. We subtract 1/2 from the 
number operators $n_{\vec{m} \sigma}=c_{\vec{m} \sigma}^{\dagger}c_{\vec{m} \sigma}$
in order for the excitations to have a maximally
symmetric form. 

In the large-$U/t$ limit, the Hamiltonian of Eq.~(\ref{eq:Hubbard})
has been shown, after a unitary transformation, to be equivalent to the Heisenberg model in the subspace
where $\langle V\rangle=0$, which is the space with exactly one electron occupying each lattice site
\cite{chao_canonical_1978,macdonald1988,oles_comment_1990}
\begin{equation}
H=J\sum_{\langle \vec i \vec j \rangle}\left(\vec{S}_{\vec{i}}\cdot\vec{S}_{\vec{j}}-\frac{1}{4}\right),\label{eq:Heisenberg}
\end{equation}
where $J$ is given by 
\begin{equation}
J=4t^{2}/U.\label{eq:JtU}
\end{equation}
The objects $\vec{S}_{\vec{m}}$ are the spin-1/2 vector operators
at lattice site $\vec{m}$. The Hamiltonian Eq.~(\ref{eq:Heisenberg})
can be derived using perturbation theory in the unitary transformation
that links the two Hamiltonians, see
Appendix \ref{app:B} for a short discussion.

The Heisenberg model has been studied extensively by the coupled-cluster
method (CCM) since the pioneering work of Ref.~\cite{roger1990thecoupledcluster}.
These studies have also been generalised to the $XXZ$ model in Ref.~\cite{bishop1991coupledcluster},
\begin{equation}
H=J\sum_{\langle \vec i \vec j \rangle}\left(S_{\vec i}^{x}S_{\vec j}^{x}+S_{\vec i}^{y}S_{\vec j}^{y}+\Delta S_{\vec i}^{z}S_{\vec j}^{z}\right)
\end{equation}
where the anisotropy parameter $\Delta$ distinguishes the various
form of the $XXZ$ model ($\Delta>0$). The CCM analysis starts from
the classical Ising limit ($\Delta\to\infty$) and includes quantum
correlations in the ground state, which of course depend on the anisotropy.
In particular, the analysis shows that the spin-spin correlations
show algebraic decay as the anisotropy decreases to a critical value
$\Delta=\Delta_{c}$. For example, on a square lattice at the critical
anisotropy the spin-wave excitations become gapless with a value of
spin-wave velocity in agreement with that of the second order spin-wave
theory of Anderson \cite{anderson1952anapproximate,kubo1952thespinwave,oguchi1960theoryof}
at the isotropic point $\Delta=1$; for the one-dimensional model,
one finds the expected value zero for the sub-lattice magnetization
at the critical anisotropy, in contrast to the divergent result from
spin-wave theory \cite{anderson1952anapproximate}. There are 
good theoretical arguments that
 $\Delta_c$ converges to 1 as we increase the order of the CCM calculation
.

In this paper, we apply a similar CCM analysis. We start from a N\'eel state  and include quantum
many-body correlations  by considering correlations caused by excitations (both charge and spin)
on top of this state. As expected, our results for the
ground-state energies of the Hubbard model of Eq.~(\ref{eq:Hubbard})
reduce to those of the spin models in the large-$U/t$ limit using
corresponding CCM truncations. As we shall discuss in more detail
below, this correspondence requires a very subtle incorporation of
the Heisenberg model results into the Hubbard model, including
the incorporation of the unitary transformation.

 Therefore, for general values
of $U/t$, we take the advantage of the results from the solution of
the $XXZ$ model with the anisotropy as a parameter. We shall show
that it makes sense to use the critical value, and directly employ
the resulting algebraic two-body spin-spin correlations at the critical
anisotropy in our study of the Hubbard model. Indeed, as we will demonstrate,
the ground-state energies are at minimum at the critical anisotropy.

\section{Coupled-Cluster Method and the Super-SUB$n$ approximation\label{sec:Coupled-Cluster-Method}}

In the normal coupled-cluster method (NCCM) we describe the ground
state of an interacting system as the exponential of a generalised
creation operator acting on a generalised vacuum state (in a more group-theoretical setting, an extremal weight state)
\cite{bishop1987thecoupledcluster,bishop1991anoverview},
\begin{equation}
|\Psi\rangle=e^{S}|\Phi_{0}\rangle.\label{eq:CCMwf}
\end{equation}
The key idea of the CCM, is that
we do not assume the bra state to be the Hermitian conjugate of the
ket state. Effectively this corresponds to using a bi-orthogonal basis, where 
the dual states are not the Hermitian conjugates of the ket states. The advantage will be that all expressions are finite
polynomials in the correlations, but the disadvantage can be that we no longer have a variational upper bound
to the energy. We use the parametrisation
\begin{equation}
\langle\tilde{\Psi}|=\langle\Phi_{0}|(1+\tilde{S})e^{-S}.\label{eq:CCMbra}
\end{equation}
The operators $S$ and $\tilde{S}$ are then expanded in generalised
(multi-particle) creation and annihilation operators, 
\begin{align*}
C_{I}\left|\Phi_{0}\right\rangle  & =0\text{ if \ensuremath{I>0}},
\end{align*}
\begin{equation}
S=\sum_{I\neq0}s_{I}C_{I}^{\dagger},\qquad\tilde{S}=\sum_{I\neq0}\tilde{s}_{I}C_{I},\label{eq:CCMops}
\end{equation}
with many-body correlation coefficients $s_{I}$ and $\tilde{s}_{I}$
to be determined. Here we conventionally choose $C_{0}$ as the identity
operator, and thus the inequality in the summations in Eq.~(\ref{eq:CCMops})
excludes a constant term.

The expansion (\ref{eq:CCMops}) together with Eqs.\ (\ref{eq:CCMwf}) and (\ref{eq:CCMbra}) can give
an in principle exact description for the ground state, by applying the
variational principle to the energy.
The ground state solution for the set of coefficients $s_{I}$ is then
given by the non-linear equations obtained from the variation of the expectation value of the Hamiltonian, $\langle H \rangle=\braOket{\tilde\Psi}{H}{\Psi}$,  with respect to 
$\tilde{s}_I$,
\begin{align}
\forall_{I>0}:\ \left\langle \Phi_{0}\right|C_{I}e^{-S}He^{S}\left|\Phi_{0}\right\rangle  & =0.\label{eq:CCMCI}\end{align}
Using this equation, and the fact that the general energy expression is linear in $\tilde{s}$, we see that $\tilde{s}$ does not contribute to the ground-state energy,
\begin{align}
E_{0} & =\left\langle \Phi_{0}\right|e^{-S}He^{S}\left|\Phi_{0}\right\rangle .\label{eq:CCME}
\end{align}
The coefficients $\tilde{s}_{I}$, sometimes called bra-state coefficients,
are thus not required for the evaluation of the ground state energy, but they
do enter the expectation value of other observables through Eq.\ (\ref{eq:CCMbra}). They can
 determined
from the linear equations, which follow from the variation of the general expression for  the energy 
with respect to $s_J$,
\begin{equation}
\sum_{I>0}\tilde{s}_{I}\left\langle \Phi_{0}\right|C_{I}e^{-S}[H,C_{J}^{\dagger}]e^{S}\left|\Phi_{0}\right\rangle =0,\label{eq:CCMSt}
\end{equation}
once we have determined the values of $s_{I}$ from Eq.~(\ref{eq:CCMCI}).

Since there are only a finite--typically small--number of possible contractions, the equation (\ref{eq:CCME}) expresses the ground-state energy in
terms of a small subset of the CCM coefficients $s_{I}$. Normally
the CCM equations (\ref{eq:CCMCI}) involve all of the coefficients, due to the presence of the operators $C_I$.
In many cases we can identify a hierarchy in these equations--that
is commonly based on the number of basic (single-particle) operators
that make up the operator $C_{I}^{\dagger}$. We then label successive
terms in this hierarchy with an index $n$, and we denote the SUB$n$
approximation as the case where we use all the creation and annihilation
operators up to level $n$ in the hierarchy. In principle we can systematically
improve on these calculations by simply increasing $n$--though the
complications increase rapidly with $n$.

\subsection{CCM for the Hubbard model}

The most comprehensive application of the coupled-cluster method to
the Hubbard model can be found in Ref.~\cite{bishop1995amicroscopic},
see also the earlier work \cite{roger1990thecoupledcluster}. As is
discussed in those papers, the common choice of reference state is
the Néel state
\begin{equation}
|\Phi_{0}\rangle=\prod_{\vec{i},\vec{j}}c_{\vec{i}\uparrow}^{\dagger}c_{\vec{j}\downarrow}^{\dagger}|0\rangle,
\end{equation}
i.e., an antiferromagnetic state where the $A$ sub-lattice is magnetised
upwards, and the $B$ one downwards--so all nearest neighbours are
in the classically optimal position of having their spins pointing
in opposite directions. For this particular choice of reference state,
it is easier to work with quasi-particle operators that are 
 the single particle creation and annihilation operators relative
to the Néel state (an extreme case of a Bogoliubov transformation)
\begin{align}
a_{\vec{i}\uparrow} & =c_{\vec{i}\uparrow}^{\dagger},\quad a_{\vec{i}\downarrow}=c_{\vec{i}\downarrow},\nonumber \\
b_{\vec{j}\uparrow} & =c_{\vec{j}\downarrow}^{\dagger},\quad b_{\vec{j}\downarrow}=c_{\vec{j}\uparrow}.
\end{align}
In terms of these new operators we have %
\footnote{Here we use the notation $n_{a}=\sum_{\sigma,\vec{i}}a_{\vec{i}\sigma}^{\dagger}a_{\vec i\sigma}$,
and similar for $n_{b}$.%
} 
\begin{align}
H & =-t\sum_{\langle\vec{i}\vec{j}\rangle}\left(b_{\vec{j}\uparrow}a_{\vec{i}\downarrow}-b_{\vec{j}\downarrow}a_{\vec{i}\uparrow}+a_{\vec{i}\downarrow}^{\dagger}b_{\vec{j}\uparrow}^{\dagger}-a_{\vec{i}\uparrow}^{\dagger}b_{\vec{j}\downarrow}^{\dagger}\right)+\nonumber \\
 & -U\sum_{\vec{i}}a_{\vec i\downarrow}^{\dagger}a_{\vec i\uparrow}^{\dagger}a_{\vec i\uparrow}a_{\vec i\downarrow}-U\sum_{\vec{j}}b_{\vec j\downarrow}^{\dagger}b_{\vec j\uparrow}^{\dagger}b_{\vec j\uparrow}b_{\vec j\downarrow}+\frac{U}{2}(n_{a}+n_{b}).\label{eq:Hubbardqp}
\end{align}
We now expand the CCM correlations in terms of powers of the creation
operators $a^{\dagger}$ and $b^{\dagger}$, the SUB$n$ expansion, as
\begin{equation}
S=\sum_{k=1}^{n}S_{k},\,\tilde{S}=\sum_{k=1}^{n}\tilde{S}_{k}.
\end{equation}
Since the total spin projection quantum number is conserved, we always
find an equal number of spin up and spin down operators in each $S_{n}$,
and thus an equal number of $a$ and $b$ operators. The lowest order term
takes the form 
\begin{align}
S_{1} & =\sum_{\vec{i}\vec{j}}s_{\vec{i}\vec{j}}\left(a_{\vec{i}\uparrow}^{\dagger} b_{\vec{j}\downarrow}^{\dagger}- a_{\vec{i}\downarrow}^{\dagger}b_{\vec{j}\uparrow}^{\dagger}\right),\label{eq:S1}
\end{align}
i.e., where the CCM operators $C_I$ are an antisymmetric combination
of one $a$ and one $b$ creation operator.

It may be interesting here to comment on the choice of antisymmetry
under spin exchange of the operator in Eq.~(\ref{eq:S1}), which is
not immediately obvious from the discussion above. It is actually
more restrictive than one would expect, but additional analysis shows
that it is an operator that adds a spin-zero pair of quasi-particles
to the Néel state. That explains why this is the correct structure: it is due to the fact
that the Hamiltonian (\ref{eq:Hubbardqp}) is actually a quasi-particle
spin $0$ operator, and the Néel state, which is the
quasi-particle vacuum state, has spin zero as well. Thus any correlated
state build upon this must also have this symmetry.  Alternatively,
this structure can be shown to be correct for the ground state due to its symmetry under
exchange of $a_{i\sigma}^{\dagger}\leftrightarrow b_{j\bar{\sigma}}^{\dagger}$ (where the bar denotes a spin-flip, $\bar{\uparrow}=\downarrow$, $\bar{\downarrow}=\uparrow$)
together with the anticommutation of the fermion operators. 

The most general $S_{2}$ operator can be decomposed in three components, 
\begin{align}
S_{2} & =\sum_{\vec{i}\vec{i}'\vec{j}\vec{j}'} \bigg(s_{\vec{i}\vec{i}'\vec{j}\vec{j}'}^{(1)}a_{\vec{i}\uparrow}^{\dagger} a_{\vec{i}'\downarrow}^{\dagger}b_{\vec{j}\downarrow}^{\dagger} b_{\vec{j}'\uparrow}^{\dagger}+\nonumber \\
 & \qquad\quad s_{\vec{i}\vec{i}'\vec{j}\vec{j}'}^{(2)} a_{\vec{i}\uparrow}^{\dagger}a_{\vec{i}'\uparrow}^{\dagger} b_{\vec{j}\downarrow}^{\dagger}b_{\vec{j}'\downarrow}^{\dagger} +s_{\vec{i}\vec{i}'\vec{j}\vec{j}'}^{(3)} a_{\vec{i}\downarrow}^{\dagger}a_{\vec{i}'\downarrow}^{\dagger} b_{\vec{j}\uparrow}^{\dagger}b_{\vec{j}'\uparrow}^{\dagger}\bigg).
\end{align}
We have a similar form for $\tilde{S}$, but now in term of annihilation
operators, 
\begin{equation}
\tilde{S}_{1}=\sum_{\vec{i}\vec{j}}\tilde{s}_{\vec{i}\vec{j}} \left(b_{\vec{j}\downarrow}a_{\vec{i}\uparrow} -b_{\vec{j}\uparrow}a_{\vec{i}\downarrow}\right),
\end{equation}
and similar for $\tilde{S}_2$.
Explicit investigation of the SUB2 truncation, retaining only $S_{1}$
and $S_{2}$, shows that the coefficients $s^{(2)}$ and $s^{(3)}$are solutions to a homogeneous linear problem, and are thus zero in
the ground state  %
\footnote{This can again be explained in terms of the \emph{quasi-particle}
spin symmetry: There is only one spin zero operator.}.
Since the ground state is translationally invariant, we find that
independent of truncation the coefficients $s$ (\ref{eq:S1}) are also translationally
invariant, $s_{\vec{i}\vec{j}}=s_{\vec{j}-\vec{i}}\equiv s_{\vec{r}}$.

Here, and in the remainder of this paper we shall use the symbol $\vec{r}$
to denote a vector pointing from a point on the $A$ sublattice to
a point on the $B$ sublattice. We shall also use the symbol $\vec{\rho}$
to denote the values of $\vec{r}$ that connect nearest neighbours.
Solving the CCM equations, we find that all $s_{\vec{\rho}}$ are the same, $s_{\vec{\rho}}=s_{1}$,
since the lattice symmetries assure that these parameters are direction
independent in the ground state.

The exact expression for the energy in the CCM approach is given by
(here $z$ is the lattice coordination number, the number of nearest neighbours of every lattice point)
\begin{equation}
E_{0}/N=t\sum_{\vec{\rho}}s_{\vec{\rho}}\equiv zts_{1},\label{eq:EnerCCM}
\end{equation}
which only depends on the value of $s_{1}$. By selecting those equations
from Eq.~(\ref{eq:CCMCI}) where $C_{I}$ consists of one $a$ and
one $b$ annihilation operator, we find the one-body equation
\begin{equation}
2t\sum_{\vec{\rho}}\left(\delta_{\vec{r}\vec{\rho}} -\sum_{\vec{r}'}s_{\vec{r}'}s_{\vec{r}-\vec{r}'+\vec{\rho}}\right) +2Us_{\vec{r}} +t\sum_{\vec{i}_{1}}\sum_{\vec{\rho}}\left(s_{\vec{i}_{1}\vec{i}_{2}\vec{i}_{1}+\vec{\rho},\vec{i}_{2}+\vec{r}}^{(1)}+s_{\vec{i}_{2}\vec{i}_{1}\vec{i}_{2}+\vec{r},\vec{i}_{1}+\vec{\rho}}^{(1)}\right)=0.\label{eq:SUB1}
\end{equation}
The second order coefficient $s^(1)$ appears in this equation due to contraction with the Hamiltonian 
in the evaluation of $e^{-S}He^S$.
This equation is exact for any SUB$n$ truncation with $n\ge2$.
Similarly, the two-body equations (obtained for $C_{I}$'s consisting
of two $a$ and two $b$ operators) will involve higher-order coefficients
as well. This leads to an infinite hierarchy of equations, which require
a closure approximation or even a truncation, in order to make the
equations tractable.

\subsubsection{SUB1 approximation}

The simplest truncation to make is the SUB1 approximation, by which
we denote a calculation where we only include the $S_{1}$ operator. 
It is quite illustrative to work through the derivation of these results in some detail to illustrate the methodology; 
for the more complicated calculations in the following sections the derivation is given in Appendix \ref{app:A}.

From Eq.~(\ref{eq:SUB1}) we find the
one-body equation
\begin{equation}
t\sum_{\vec{\rho}}\left(\delta_{\vec{r}\vec{\rho}}-
\sum_{\vec{r}'}s_{\vec{r}'}s_{\vec{r}-\vec{r}'+\vec{\rho}}\right)+Us_{\vec{r}}=0.
\label{eq:obd}
\end{equation}
This can be solved by a sublattice Fourier transform, see, e.g., \cite{bishop1991coupledcluster},
by writing 
\begin{align}
s_{\vec{q}} & =\sum_{\vec{r}}e^{i\vec{q}\cdot\vec{r}}s_{\vec{r}},\\
s_{\vec{r}} & =\frac{1}{|\mathcal{A}|}\int_{_{\mathcal{A}}}e^{-i\vec{q}\cdot\vec{r}}s_{\vec{q}}\, d\vec{q},
\end{align}
and, when required (note the complex conjugate Fourier transform),
\begin{equation}
\tilde{s}_{\vec{r}}=\frac{1}{|\mathcal{A}|}\int_{_{\mathcal{A}}}e^{i\vec{q}\cdot\vec{r}}\tilde{s}_{\vec{q}}\, d\vec{q}.
\end{equation}
Here $\mathcal{A}$ denotes the first Brillouin zone (FBZ) of the
$B$ sub-lattice, and $|\mathcal{A}|$ is its area. Using the sublattice
Fourier transform gives the equation 
\begin{equation}
tz\left(\gamma_{\vec{q}}-\gamma_{-\vec{q}}s_{\vec{q}}^{2}\right)+Us_{\vec{q}}=0,\label{eq:SLFT}
\end{equation}
where 
\begin{equation}
\gamma_{\vec{q}}\equiv\frac{1}{z}\sum_{\vec{\rho}}e^{i\vec{q}\cdot\vec{\rho}}.
\end{equation}
On a general bi-partite lattice, $\gamma_{-\vec{q}}=\gamma_{\vec{q}}^{*}$.
Equation (\ref{eq:SLFT}) can now be solved as a quadratic equation.
Choosing the physical root, one finds 
\begin{equation}
s_{1}=\frac{1}{k}\frac{1}{|\mathcal{A}|}\int_{_{\mathcal{A}}}\left(1-\sqrt{1+k^{2}|\gamma_{\vec{q}}|^{2}}\right)d\vec{q},
\end{equation}
where $k$ is the coordination-weighted ratio of coupling constants,
\begin{equation}
k=2zt/U,
\end{equation}
as in Eq.~(18) of Ref.~\cite{bishop1995amicroscopic}.

\subsubsection{SUB2 on-site approximation}

In the SUB2 approximation, where we include also the $S_{2}$ operator,
the energy equation (\ref{eq:EnerCCM}) is unchanged, but we need
to include the exact one-body CCM equation (\ref{eq:SUB1}) and make
an approximation to the two-body one,

\begin{align}
-U\left[\left(s_{\vec{i}_{1}\vec{i}_{2}\vec{j}_{1}\vec{j}_{1}}^{(1)}+s_{\vec{j}_{1}-\vec{i}_{2}}s_{\vec{j}_{1}-\vec{i}_{1}}\right)\delta_{\vec{j}_{1}\vec{j}_{2}}+\left(s_{\vec{i}_{1}\vec{i}_{1}\vec{j}_{1}\vec{j}_{2}}^{(1)}+s_{\vec{j}_{1}-\vec{i}_{1}}s_{\vec{j}_{2}-\vec{i}_{1}}\right)\delta_{\vec{i}_{1}\vec{i}_{2}}-2s_{\vec{i}_{1}\vec{i}_{2}\vec{j}_{1}\vec{j}_{2}}^{(1)}\right]\nonumber \\
-t\sum_{\vec{\rho}}\left[\sum_{\vec{i}_{3}}\left(s_{\vec{i}_{3}+\vec{\rho}-\vec{i}_{1}}s_{\vec{i}_{3}\vec{i}_{2}\vec{j}_{1}\vec{j}_{2}}^{(1)}+s_{\vec{i}_{3}+\vec{\rho}-\vec{i}_{2}}s_{\vec{i}_{1}\vec{i}_{3}\vec{j}_{1}\vec{j}_{2}}^{(1)}\right)+\sum_{\vec{j}_{3}}\left(s_{\vec{j}_{1}-\vec{j}_{3}+\vec{\rho}}s_{\vec{i}_{1}\vec{i}_{2}\vec{j}_{3}\vec{j}_{2}}^{(1)}+s_{\vec{j}_{2}-\vec{j}_{3}+\vec{\rho}}s_{\vec{i}_{1}\vec{i}_{2}\vec{j}_{1}\vec{j}_{3}}^{(1)}\right)\right] & =0.\label{eq:tbeq}
\end{align}
The lattice symmetries require that for the ground state $s^{(1)}$ is symmetric
under interchange of the $i$ and $j$ indices. The two-body equation
(\ref{eq:tbeq}) is very hard to solve, as it contains
objects with four independent indices; a simple first approximation
is to choose a subset of coefficients, those with $i_{1}=i_{2}$ and
$j_{1}=j_{2}$, and require those to be the only non-zero ones. In
this on-site (OS) approximation, we thus have 
\begin{equation}
s_{\vec{i}_{1}\vec{i}_{2}\vec{j}_{1}\vec{j}_{2}}^{(1)}=\delta_{\vec{i}_{1}\vec{i}_{2}}\delta_{\vec{j}_{1}\vec{j_{2}}}s_{\vec{j}_{1}-\vec{i}_{1}}^{(1)},
\end{equation}
and a similar relation for the coefficients $\tilde{s}^{(1)}$. 

This makes it straightforward to derive the CCM equations, see Appendix \ref{App:A-SUB2OS} for details.

\subsubsection{Super-SUB1 Approximation}

As we shall show below, the solution of the truncated CCM equations
in the OS approximation only gives a slight improvement on the simple
SUB1 truncation. We believe that this is due to the fact that this approximation
does not contain some important correlations. In other words, we may
need to consider the SUB3 truncation for the Hubbard model. This may come
as a surprise since for the Heisenberg model the SUB2
scheme is highly accurate. Due to the fact that we need to perform a unitary transformation
to link the two models, 
in the Hubbard model, we can only describe similar correlations in
the SUB3 approximation. This would be a very challenging calculation,
and therefore we investigate an alternative closure approximation
which includes the most important effects of the SUB3 truncation, but does not require
a direct evaluation. We take advantage of the fact that the exact
one-body equation Eq.~(\ref{eq:SUB1}) only contains $S_{1}$ and
$S_{2}$ coefficients, and we take the SUB2 coefficients $s_{\vec{r}}^{(1)}$
and $\tilde{s}_{\vec{r}}^{(1)}$ from a related calculation. A natural
choice would be the the CCM solution of the Heisenberg model, but
as discussed before
we shall use the more general spin-$1/2$ $XXZ$ model. Thus we choose
$s_{\vec{r}}^{(1)}=\alpha_{\vec{r}}^{\Delta}$ and $\tilde{s}_{\vec{r}}^{(1)}=\tilde{\alpha}_{\vec{r}}^{\Delta}$,
where we use $\alpha_{\vec{r}}^{\Delta}$ and $\tilde{\alpha}_{\vec{r}}^{\Delta}$
to refer to the ket and bra SUB2 coefficients for the
$XXZ$ model with anisotropy factor $\Delta$ \cite{bishop1998thecoupled}.
Strictly speaking, the parameter $\Delta$ should be $1$, since, as stated before,
the Hubbard model goes to the $\Delta=1$ Heisenberg model
in the large $U/t$ limit. We prefer to find the optimal
choice of $\Delta$ for finite $U/t$. We shall show that the energy
is minimal for the critical value of $\Delta$, where the CCM coefficients
generate power-law decay of the correlation functions \cite{bishop1998thecoupled}.
This critical behaviour is crucial in describing the one-dimensional model,
and we shall argue that the critical value of $\Delta$ is the optimal choice.
Explicit expressions for the CCM parameters are given in Appendix \ref{App:A-Super}.

\subsubsection{Link to the Heisenberg model}

If we want to exploit the link to the Heisenberg model more fully,
we first need to investigate the behaviour of our results in the limit
$U\rightarrow\infty.$ It is straightforward to show that in the SUB2
on-site approximation, the $s^{(1)}$ coefficients collapse to the
double-flip coefficients of the SUB2-1 approximation for the $XXZ$
model at $\Delta=1$. Here one retains the full set of SUB1 coefficients
and only the nearest neighbour SUB2 coefficient $s_{1}^{(1)}$. One
finds that, for any bipartite lattice with coordination number $z$,
\begin{align}
s_{1}^{(1)}\Bigr|_{U\rightarrow\infty}=\frac{1}{(2z-1)}.
\end{align}
The non-zero limit of the nearest-neighbour coefficients reflects
the fact that the Néel state is not the quantum ground state in the
large $U/t$ limit. This approximation also reproduces an approximation
to the ground-state energy of the Heisenberg model. We find, neglecting
the constant term, 
\begin{align}
\frac{E_{0}}{N}\Bigr|_{U\rightarrow\infty}=-z\frac{t^{2}}{U}\Big(1+s_{1}^{(1)}\Bigr).
\end{align}
 If we compare this to the $XXZ$-model ground-state energy in the
SUB2-1 approximation, 
\begin{align}
\frac{E_{0}}{N}=-J\frac{z}{8}\Bigl(1+2\alpha_{1}^{\Delta}\Bigr)-J\frac{z}{8},
\end{align}
 and use the relation (\ref{eq:JtU}), we see that these two indeed
agree.

\subsection{Excitation energies\label{sub:Excitation-energies}}

There are two equivalent ways to derive the excitation energy from
the CCM. The first is the bi-variational method, where we derive the
excitation energies from the variations about stable equilibrium in
the time-dependent variational method (sometimes called ``generalised
RPA'' or ``Harmonic Approximation''),
\begin{equation}
\delta \braOket{\tilde{\Psi}}{i\partial_t-H}{\Psi},
=0
\end{equation}
where we use  the CCM states (\ref{eq:CCMwf}) and (\ref{eq:CCMbra}), but now
with all the CCM coefficients depending on time.
This shows the fundamental
connectivity of the excitations to the ground state calculation. The
disadvantage of this method is that we need to write the CCM variational
functional ignoring the symmetries of the ground state, since the
excited states do not share the symmetries of the ground state.

There is an alternative but completely equivalent method due to Emrich
\cite{emrich1984electron,emrich1981anextension2,emrich1981anextension}
based on a linearisation of the time-dependent Schrödinger equation
in terms of the excitation operator  $X=\sum_{J}\chi_{J}C_{J}^{\prime\dagger}$,
which acts on the correlated CCM state to give the excited state
$X\ket{\Psi}$.
From the Schrödinger equation for this state, using the fact that $S$ and $X$ commute, we derive
\begin{equation}
e^{-S} H e^S  X \ket{\Phi_0}=EX \ket{\Phi_0},
\end{equation}
we can, by using projection on the states $\bra{\Phi_0}C^\prime_I$, subtracting the ground state energy, and using Eqs.~(\ref{eq:CCMwf},\ref{eq:CCMbra},\ref{eq:CCMops}),
obtain the equations
\begin{align}
\sum_{J}\left\langle \Phi_{0}\right|C_{I}^{\prime}e^{-S}[H,C_{J}^{\prime\dagger}]e^{S}\left|\Phi_{0}\right\rangle \chi_{J} & =\omega\chi_{I},
\end{align}
This is a linear eigenvalue problem for the excitation energies $\omega$. One should keep in
mind that \emph{in principle} we are not guaranteed that the eigenvalues
are real, since CCM does not guarantee hermiticity--the fact that
all physical eigenvalues have to be real can be used an important
check on the quality of the approximations made to obtain the results.
The reason we label the operators $C$ by a prime is that we usually
consider creation operators that do not have the symmetry of the ground
state, and they are thus not the same as the operators $C$ that occur
in the ground-state calculation.

In this paper we shall consider both charge excitations and spin-flip
modes. We shall label the energy spectrum by the ``good quantum numbers'',
particle number $n$ and total quasi-particle spin $S_{\text{tot}}$,
and spin projection $S_{\text{tot}z}$ 
\begin{equation}
E=E(n,S_{\text{tot}},S_{\text{tot}z}).
\end{equation}

\subsubsection{Charge excitations\label{sub:Charge-gap}}

We first look at single-particle and single-hole (charge) excitations.
We associate the operators $X^{h,p}$ with coefficients $\chi_{I}^{h,p}$,
 where the set of indices $\{I\}$ differs for electrons/particles
($p$) and holes ($h$). In the
simplest approximation, we consider excitation operators that contain
only a single quasi-particle operator, 
\begin{align}
X^{h} & =\sum_{\vec{i}}^{N/2}\chi_{\vec{i}}^{h}\, a_{\vec{i}\uparrow}^{\dagger},\label{eq:Spincharge4}\\
X^{p} & =\sum_{\vec{i}}^{N/2}\chi_{\vec{i}}^{p}\, a_{\vec{i}\downarrow}^{\dagger},
\end{align}
 The energy or both particle and hole states are identical, and are
the same for the SUB1 and the super-SUB1 approximations. They are
given by 
\begin{align}
\omega_{\vec{q}}^{c}=-zts_{\vec{q}}\,\gamma_{-\vec{q}}+\frac{1}{2}U.\label{eq:Spincharge6}
\end{align}
By substituting Eq.~(\ref{eq:sq}) into Eq.~(\ref{eq:Spincharge6})
we get the explicit form
\begin{align}
\omega_{\vec{q}}^{c}=\frac{U}{2}\sqrt{1+k^{2}(1+\alpha_{1}^{\Delta})\,|\gamma_{\vec{q}}|^{2}}.
\end{align}
Again, at $\alpha_{1}^{\Delta}=0$ the super-SUB1 solution collapses
to the SUB1 one.

\subsubsection{Spin-flip excitations\label{sub:Spin-excitations}}

The spin-excitation equation is obtained  within the NCCM framework
by using spin-flip operators for $C_{J}^{\prime\dagger}$. These generate
states with a non-zero total spin $S_{\text{tot}}$ without affecting
the total number of electrons, 
\begin{align}
X^{s}=\sum_{I}\chi_{I}^{s}\,\hat{C}_{I}^{\prime\dagger}.
\end{align}
 The spin-excitation energy is the difference between the energy of
the spin-flipped state, $E(N,S_{\text{tot}}\neq0)$, and the ground-state
energy $E_{0}(N,S_{\text{tot}}=0)$ at half filling, 
\begin{align}
\omega^{s}=E(N,S_{\text{tot}},S_{\text{tot}z})-E_{0}(N,0,0).
\end{align}
In this paper we shall consider the case of pure spin-flip, 
\begin{align}
\langle\Phi_{0}|a_{\vec{i}^{\prime}\downarrow}a_{\vec{i}\uparrow}e^{-S}[H,X^{s}]e^{S}|\varphi_{0}\rangle=\omega^{s}\chi_{\vec{i},\vec{i}^{\prime}}^{s},
\end{align}
 where 
\begin{align}
\omega^{s}=E(N,1,-1)-E_{0}(N,0,0),
\end{align}
 and the single spin-flip operator is defined as 
\begin{align}
X^{s} & =\sum_{\vec{i}_{1},\vec{i}_{2}}^{N/2}\chi_{\vec{i}_{1},\vec{i}_{2}}^{s}\, a_{\vec{i}_{1}\uparrow}^{\dagger}a_{\vec{i}_{2}\downarrow}^{\dagger},\label{eq:spinflipop}
\end{align}
where $\chi_{\vec{i}_1,\vec{i})2}$ are the excitation correlation
coefficients, which as the indices show both act on the A sublattice. The operator $X^S$ removes
a spin-up electron from the A sublattice, and adds an electron with the opposite
spin projection elsewhere on the same lattice.

The spin-flip equation in both SUB1 and super-SUB1 approximations
reduces to 
\begin{align}
-t\,\sum_{\langle\vec{i},\vec{j}\rangle}^{N/2}\Bigl(\chi_{\vec{i,}\vec{i}_{1}}^{s}\, s_{\vec{i}_{2},\vec{j}}+\chi_{\vec{i}_{2},\vec{i}}^{s}\, s_{\vec{i}_{1},\vec{j}}\Bigr)+U\,\chi_{\vec{i}_{1},\vec{i}_{2}}^{s}\Bigl(1-\delta_{\vec{i}_{1},\vec{i}_{2}}\Bigr)=\omega^{s}\,\chi_{\vec{i}_{1},\vec{i}_{2}}^{s}.\label{eq:spingap5}
\end{align}
 As is common, see e.g.\ Ref.~\cite{bishop1991coupledcluster}, a sublattice
plane wave solution is considered for the solution of Eq.~(\ref{eq:spingap5}),
\begin{align}
\chi_{\vec{i},\vec{i}'}^{s}=\frac{1}{|\mathcal{A}|^{2}}\int  d\vec{q}\int d\vec{q'}\,\chi_{\vec{q},\vec{q}'}^{s}\, e^{-\text{i}\vec{q}\cdot\vec{i}}\, e^{-\text{i}\vec{q}^{\prime}\cdot\vec{i}^{\prime}},\label{eq:spingapFT}
\end{align}
where both $\vec{q}$ and $\vec{q}'$ are defined on the Brillouin
zone of the A sublattice. This leads to the simple eigenvalue problem
\begin{equation}
\left(\omega_{\vec{q}_{1}}^{c}/t+\omega_{\vec{q}_{2}}^{c}/t\right)\chi_{\vec{q}_{1}\vec{q}_{2}}-\frac{U}{t}\frac{2}{N}\sum_{\vec{q}_{1}^{\prime}\vec{q}_{2}^{\prime}}\delta_{\vec{q}_{1}^{\prime}+\vec{q}_{2}^{\prime},\vec{q}_{1}+\vec{q}_{2}}^{\text{latt}}\chi_{\vec{q}_{1}^{\prime}\vec{q}_{2}^{\prime}}=\omega^s/t\,\chi_{\vec{q}_{1}\vec{q}_{2}},\label{eq:magnon}
\end{equation}
where $\omega_{\vec{q}}^{c}$ is the energy of the charge excitations
(\ref{eq:Spincharge6}), effectively the single particle-contribution to the excitation energy,
and the interaction term contains the lattice delta $\delta^{\text{latt}}$, which has
 ``Umklapp'' equivalence, i.e., vectors are taken equal
after being transformed back into the first Brillouin zone. It is
thus  natural to label these excitations by their total momentum 
$\vec{Q}=\vec{q}_{1}+\vec{q}_{2}$,
transformed back into the FBZ. Since the diagonal matrix $\left(\omega_{\vec{q}_{1}}^{c}+\omega_{\vec{q}_{2}}^{c}\right)\delta_{\vec{q}_{1}\vec{q}_{1}^{\prime}}\delta_{\vec{q}_{2}\vec{q}_{2}^{\prime}}$
does not commute with the matrix $\delta_{\vec{q}_{1}^{\prime}+\vec{q}_{2}^{\prime},\vec{q}_{1}+\vec{q}_{2}}^{\text{latt}}$,
this is actually an interesting and non-trivial eigenvalue problem.

Fortunately, it is simple to analyse the large $U/t$ limit: Here $\omega^{c}/t\rightarrow \frac {1}{2}U/t$,
so the first term in (\ref{eq:magnon}) becomes $U/t$ times the identity matrix,
and now commutes with the second term.
The second term has a block-diagonal form: each block (for fixed $\vec{Q}$)
 in
the matrix has dimension $N/2$ by $N/2$. Within each block this matrix has one
eigenvalue $-U/t$ and the remaining $N/2-1$ eigenvalues are $0$.
Within these blocks the eigenvalues $\omega^{s}/t$ are thus $N/2-1$ times $U/t$, and one
eigenvalue zero,  for every value of $\vec{Q}$. 
This zero eigenvalue has
eigenvector $\sqrt{\frac{2}{N}}(1,1,\ldots,1)$. 
Such a fully-delocalised eigenstate in Fourier space corresponds
to a local state in coordinate space ($\vec{i}_{1}=\vec{i}_{2}$ in
Eq.~(\ref{eq:spingap5})), which is the usual local (on-site) spin
flip excitation that describes the magnon states in the Heisenberg model.

Following a similar analysis for the $S_{\text{tot}}=1,S_{\text{tot}z}=1$
state we find exactly the same excitation spectrum. The third member
of the multiplet, the states with $S_{\text{tot}}=1$ and $S_{\text{tot}z}=0$
have a noninteracting spectrum, $\omega=\omega_{\vec{q}_{1}}^{c}+\omega_{\vec{q}_{2}}^{c}$,
and are thus of little interest at this level of approximation.

\subsubsection{Link to the Heisenberg model and improved spectra \label{sec:LLHeis}}
The result has a clear link to the Heisenberg model. 
At large $U/t$ we find a clear separation between a single state a low energy, and a continuum at much higher energy. The low energy state is in the space with  $\langle V \rangle=0$, which is isomorphic to the space of spin states \cite{macdonald1988}. This can thus be interpreted as the spin-wave excitation of the Heisenberg model. The high energy continuum states 
have multiple occupation on a single site, and thus occur at a much higher energy.

In the large $U/t$ limit we can again use a perturbation argument to
find the energy of the lowest state; we find that the energy of the
lowest-energy local spin-flip excitations goes like 
\begin{align}
\omega^s/t\quad & \underset{U/t\rightarrow\infty}{\rightarrow}U/t\frac{1}{2}k^{2}(1+\alpha_{1}^{\Delta})\,\frac{2}{N}\sum_{\vec{q}}|\gamma_{\vec{q}}|^{2}\nonumber \\
 & =\frac{t}{U}2z(1+\alpha_{1}^{\Delta}),\label{eq:ExLargeU}
\end{align}
which is a flat (momentum-independent) energy spectrum with a magnitude
equal to the amplitude of the spin-wave spectrum as found in the CCM approximation for the Heisenberg model \cite{bishop1991coupledcluster}.

As discussed in appendix \ref{app:B}, a more detailed analysis shows that the
only difference between this answer and the spin-wave spectrum found
in Ref.\ \cite{bishop1991coupledcluster} is the additional term
proportional to $s^{(1)}$ in the excitation energy in this
reference. Using the fact that the correspondence between Hubbard and
Heisenberg models involves both a unitary transformation and a
perturbation expansion, the simplest way to take the additional
contribution into account is just to add this term into our equation,
in the spirit of the super-SUB1 approximation for the ground state.
Thus, in coordinate space, we have to solve
\begin{align}
-\,\sum_{\langle\vec{i},\vec{j}\rangle}^{N/2}\Bigl(\chi_{\vec{i,}\vec{i}_{1}}^{s}\, s_{\vec{i}_{2},\vec{j}}+\chi_{\vec{i}_{2},\vec{i}}^{s}\, s_{\vec{i}_{1},\vec{j}}\Bigr)
-2\frac{t}{U}\sum_{\vec r,\vec\rho}s^{(1)}_{\vec{r}} \chi_{{\vec i}_1,{\vec i}_1} \delta_{\vec{i}_{1},\vec r-\vec \rho}
+\frac{U}{t}\,\chi_{\vec{i}_{1},\vec{i}_{2}}^{s}\Bigl(1-\delta_{\vec{i}_{1},\vec{i}_{2}}\Bigr)=\frac{\omega^{s}}{t}\,\chi_{\vec{i}_{1},\vec{i}_{2}}^{s}.\label{eq:spingap5a}
\end{align}
As discussed in more detail in appendix \ref{app:B}, 
as an expansion to first order in $t/U$ this expression is strictly speaking only valid for small $t/U$.
It seems a reasonable approximation for intermediate $U$, but it
definitely fails when $U$ is close to zero. As we shall see later, one
of the issues with this approximation is that the excitation energies go
below zero for small $U$.

The effect of this additional term is most easily written in Fourier
space, where the only modification to the results above is an additional $\vec Q$-dependent shift
$-2z(t^2/U)s^{(1)}(\vec Q) \gamma(\vec Q)$ of the energy of each mode.
In the large $U$ limit, the low-energy spectrum thus collapses to
\[\omega^s/t=\frac{t}{U}2z\left(1+\alpha_1^\Delta-\alpha_{\vec Q}^\Delta \gamma(\vec Q)\right) \]
which agrees with the known result for the Heisenberg model  \cite{bishop1991coupledcluster}.

\section{Results\label{sec:Results}}

\subsection{Ground state and sublattice magnetisation\label{sub:Results:gs}}

\begin{figure}
\begin{centering}
\subfloat[][The ground state energy.]{\includegraphics[width=7cm]{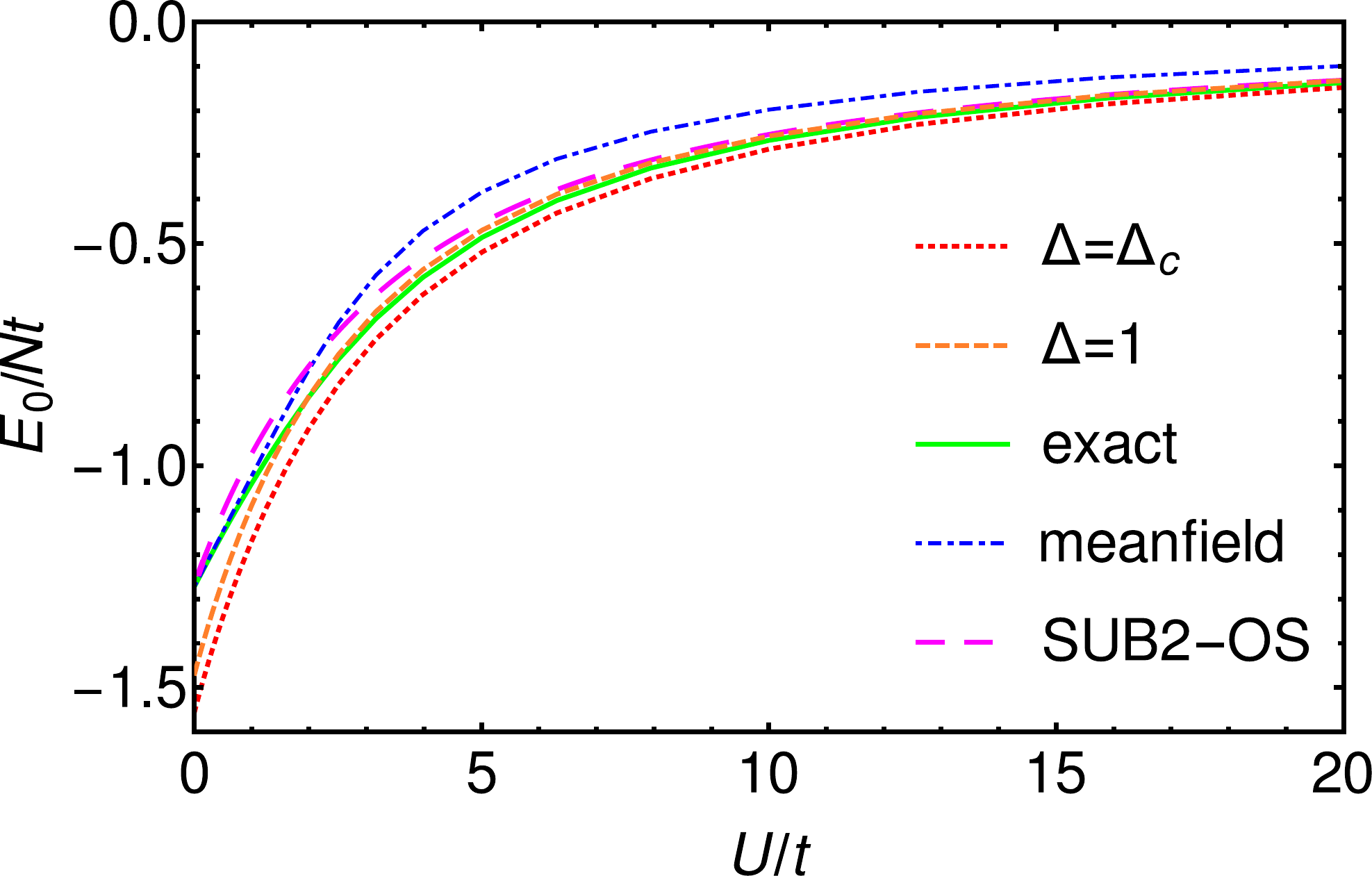}}
\subfloat[][The sub-lattice magnetisation.]{\includegraphics[width=7cm]{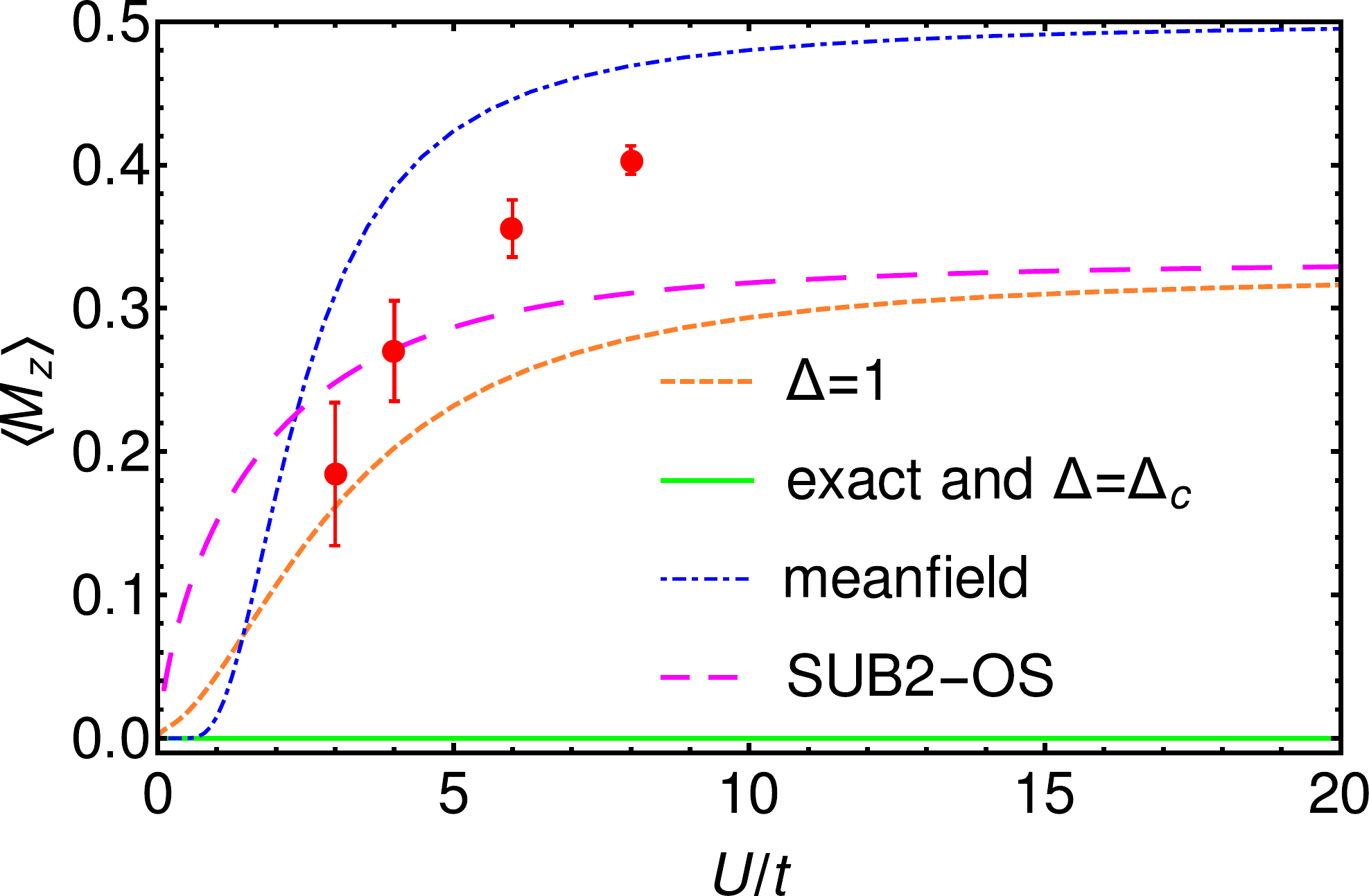}}
\end{centering}
\caption{Ground state energy and magnetic order parameter of the 1D Hubbard model in the super-SUB1
approximation for $\Delta=\Delta_c$ and $\Delta=1$ compared to the exact result.
We also show the results of mean-field theory and the CCM SUB2-OS
approximation.
The orange points (with error bars) show the Monte Carlo data from Ref.~\cite{yokoyama1987variational}.
\label{fig:Hub1d}}
\end{figure}

We first look at the ground state energy and the magnetic order parameter (the
sub-lattice magnetisation) for the Hubbard model. We start with the
exactly solvable one-dimensional model. We compare the exact result to the
super-SUB1 calculations, for the critical value of $\Delta$
($\Delta_{c}\approx 0.372755$ in 1 dimension), the SUB2-OS approximation and the mean-field results in
Figs.~\ref{fig:Hub1d}.  The calculation for the critical $\Delta$
gives the lowest energy results. These are actually below the exact
results for all values of $U/t$, but the difference is larger for small $U/t$.
For the super-SUB1 calculation for the
critical value of $\Delta$, we find the correct value of zero for the
sub-lattice magnetisation. The energy does not converge to the exact value for $U/t\rightarrow\infty$, 
that would require $\Delta=1$.
  On the other hand, both the mean-field and
SUB2-OS approximations converge to the exact result for the energy and
sub-lattice magnetisation in the limit $U/t\downarrow0 $, but they
produce poor results for the ground-state energy and the order
parameter for even moderate values of $U/t$. The fact that the
magnetisation is described correctly is a simple effect of the
algebraic nature of the correlations for $\Delta=\Delta_{c}$, 
and gives us substantial confidence in applying the same approximations
for 2D models. Henceforth we shall only look at the critical value of $\Delta$.

\begin{figure}
\begin{centering}
\subfloat[][The ground state energy.]{\includegraphics[width=7cm]{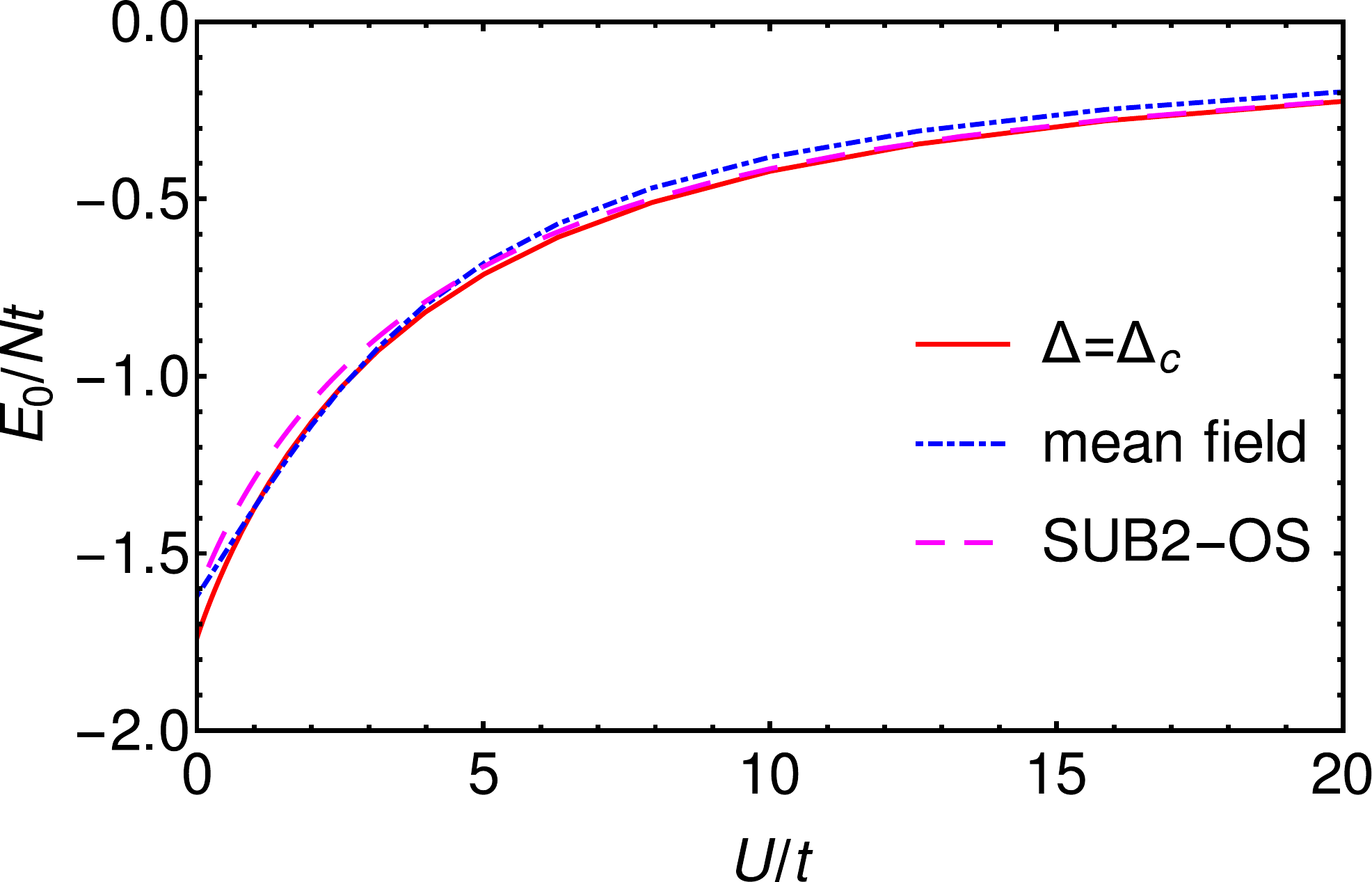}}
\subfloat[][The sub-lattice magnetisation.]{\includegraphics[width=7cm]{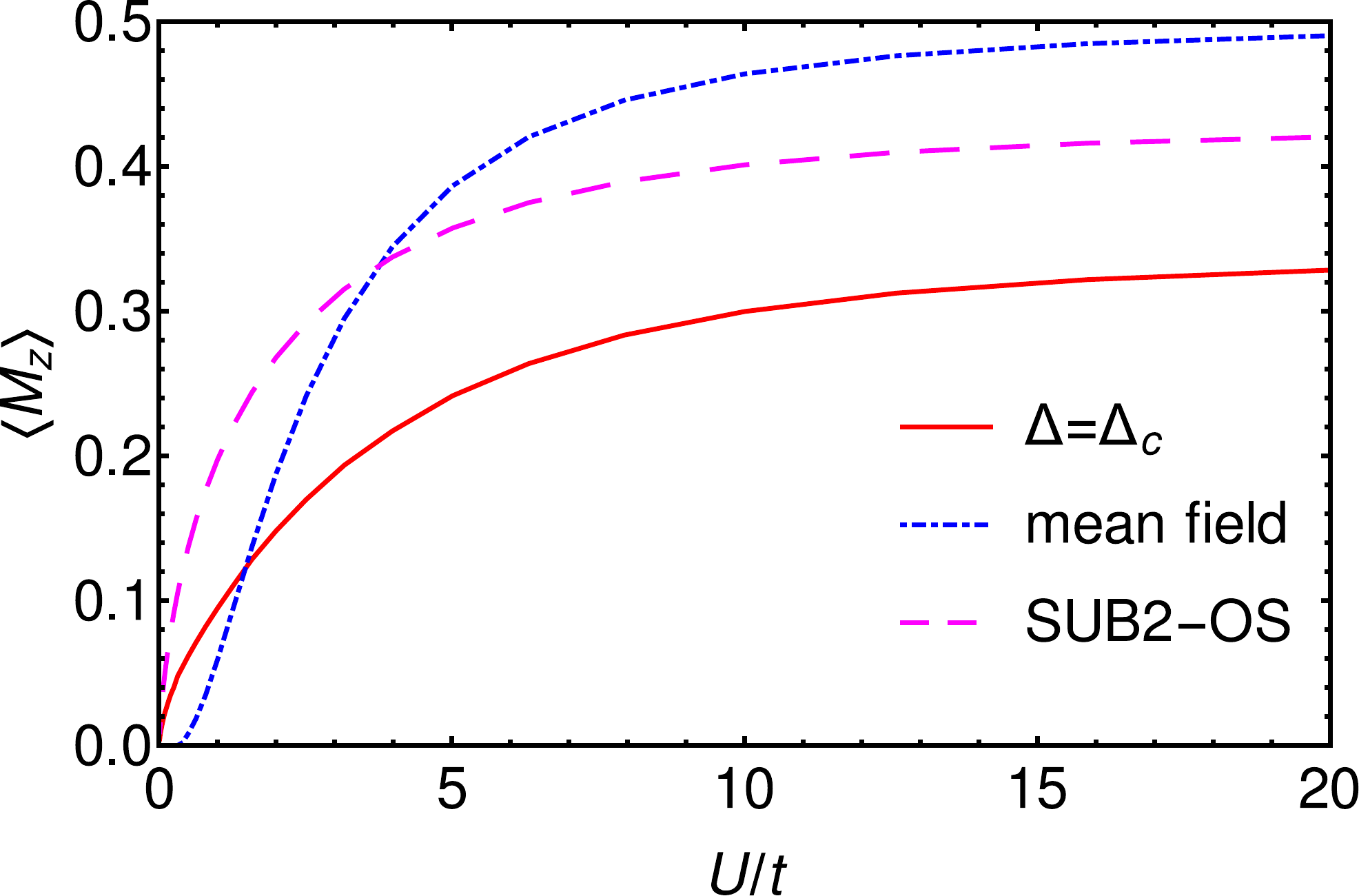}}
\end{centering}

\caption{Ground state energy and magnetic order parameter for the 2D Hubbard model on a square
lattice in the super-SUB1 approximation for  $\Delta=\Delta_c$ 
compared to mean field theory and the SUB2-OS calculation.\label{fig:Hub2dsq}}
\end{figure}

\begin{figure}
\begin{centering}
\subfloat[The ground state energy.]{\begin{centering}
\includegraphics[width=7cm]{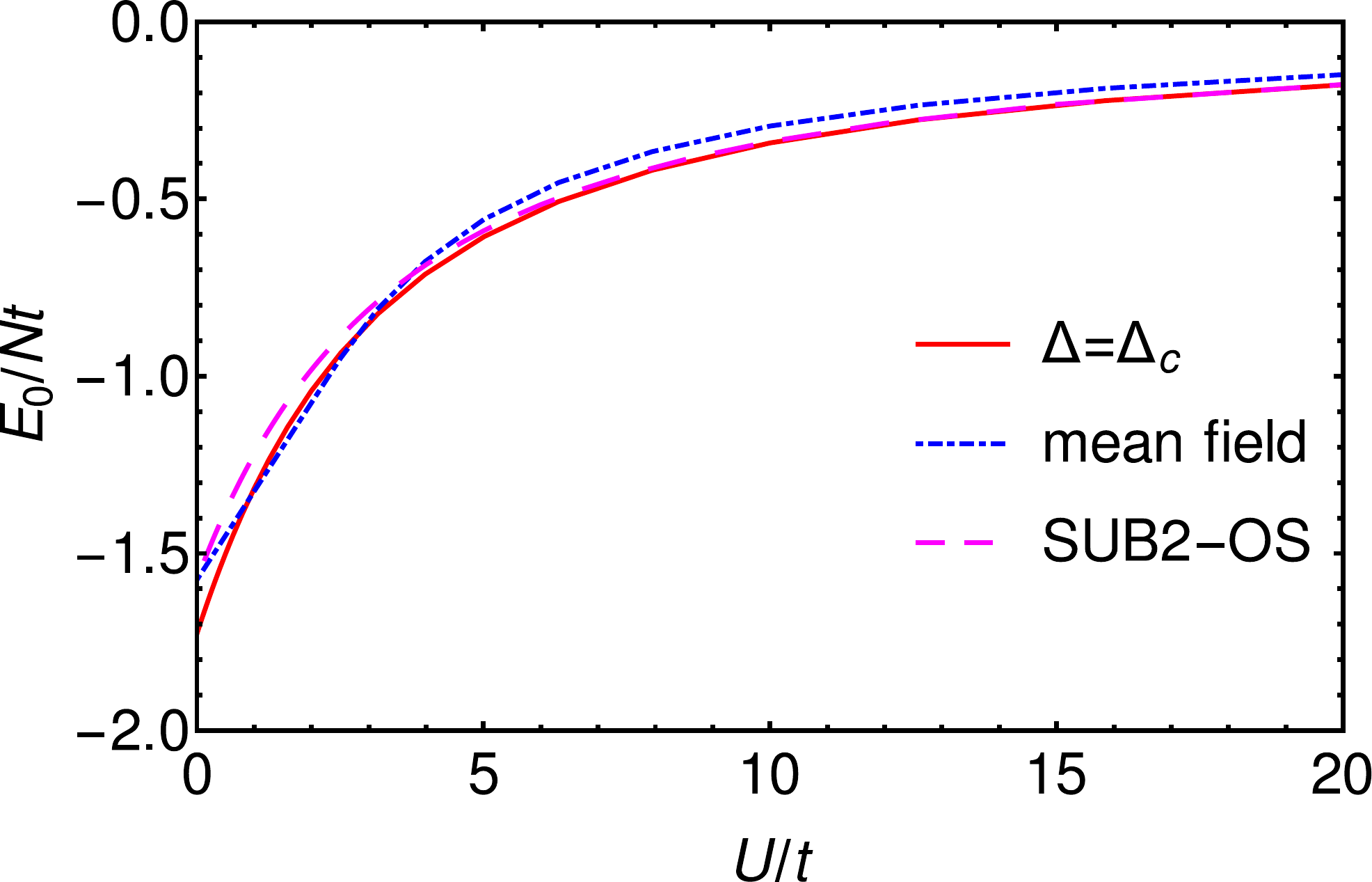}
\par\end{centering}

}\subfloat[The sub-lattice magnetisation.\label{fig:MZhx} ]{\begin{centering}
\includegraphics[width=7cm]{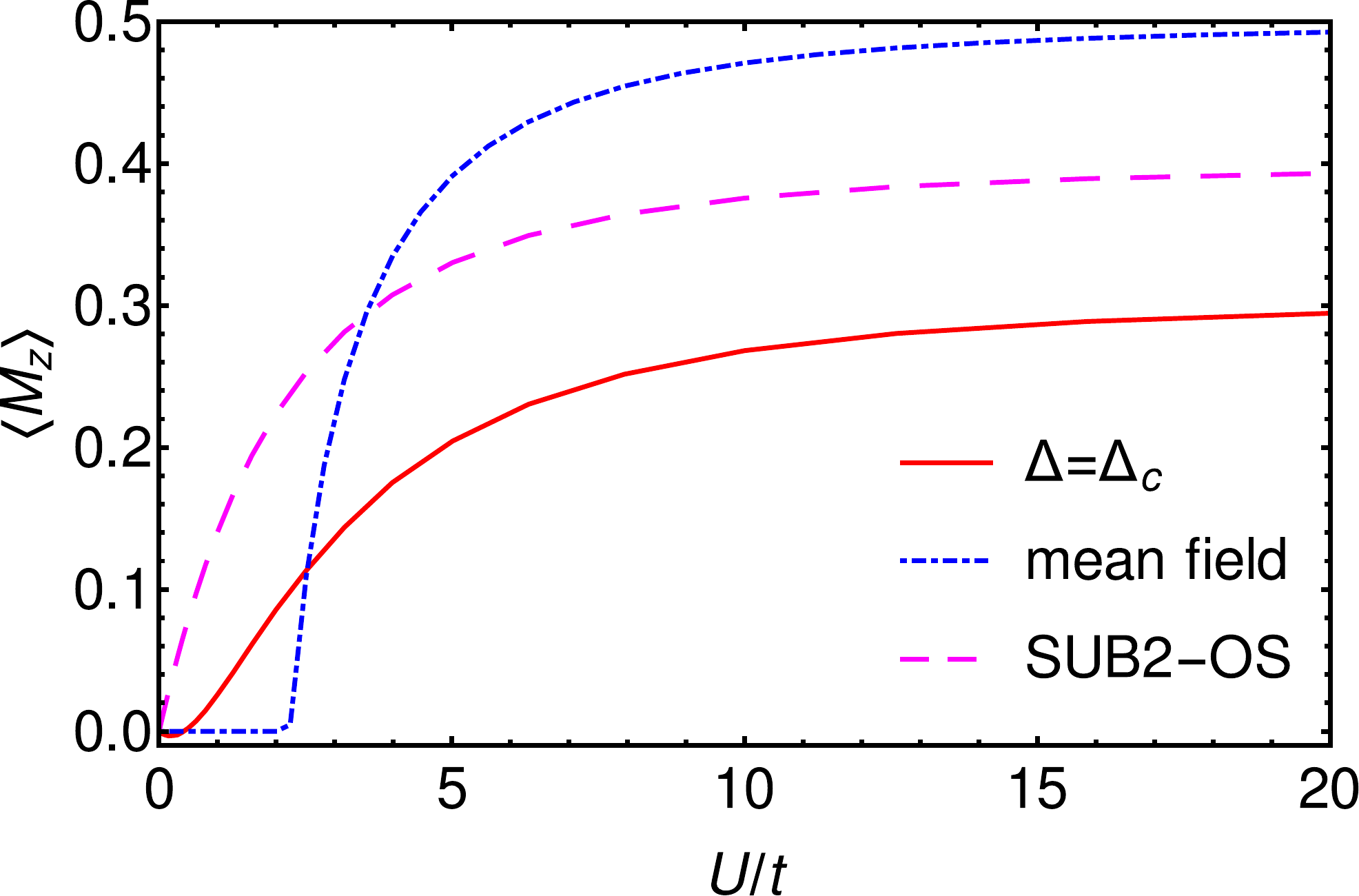}
\par\end{centering}

}
\par\end{centering}

\caption{Ground state energy and magnetic order parameter for the 2D Hubbard model on a honeycomb
lattice in the super-SUB1 approximation  for $\Delta=\Delta_c$ 
compared to mean field theory and the SUB2-OS calculation. \label{fig:Hub2dhx}}
\end{figure}

Our results for the 2D models are shown in 
Figs.~\ref{fig:Hub2dsq} and \ref{fig:Hub2dhx}. The ground
state energy in the super-SUB1 approximation shows only a weak
dependence on $\Delta$, which is why we only show the critical value results.
The values of $\Delta_{c}$
($0.7985$ for the square and $0.709826$ for the hexagonal lattice 
\footnote{The value $\Delta_c=0.709826$ for the hexagonal lattice
  disagrees with that quoted in Ref.~\cite{bishop1998thecoupled}--on
  further analysis it appears that the numerical approach applied in
  that reference is unstable for divergent integrands, as we encounter
  at the critical point.}) are rather close to $1$, so that in those
figures we only probe a small range of parameters, which explains the
similarity of the energies.  Again, using $\Delta=\Delta_{c}$ gives
the lowest ground state energy, and leads to a substantial reduction
in the sub-lattice magnetisation for large $U/t$ which is likely to be
relevant and correct, as in the 1D case. Note that the point where the
magnetisaton goes through zero, is well outside the range of validity of the
super-SUB1 approximation.

\begin{figure}
\begin{centering}
\includegraphics[width=9cm]{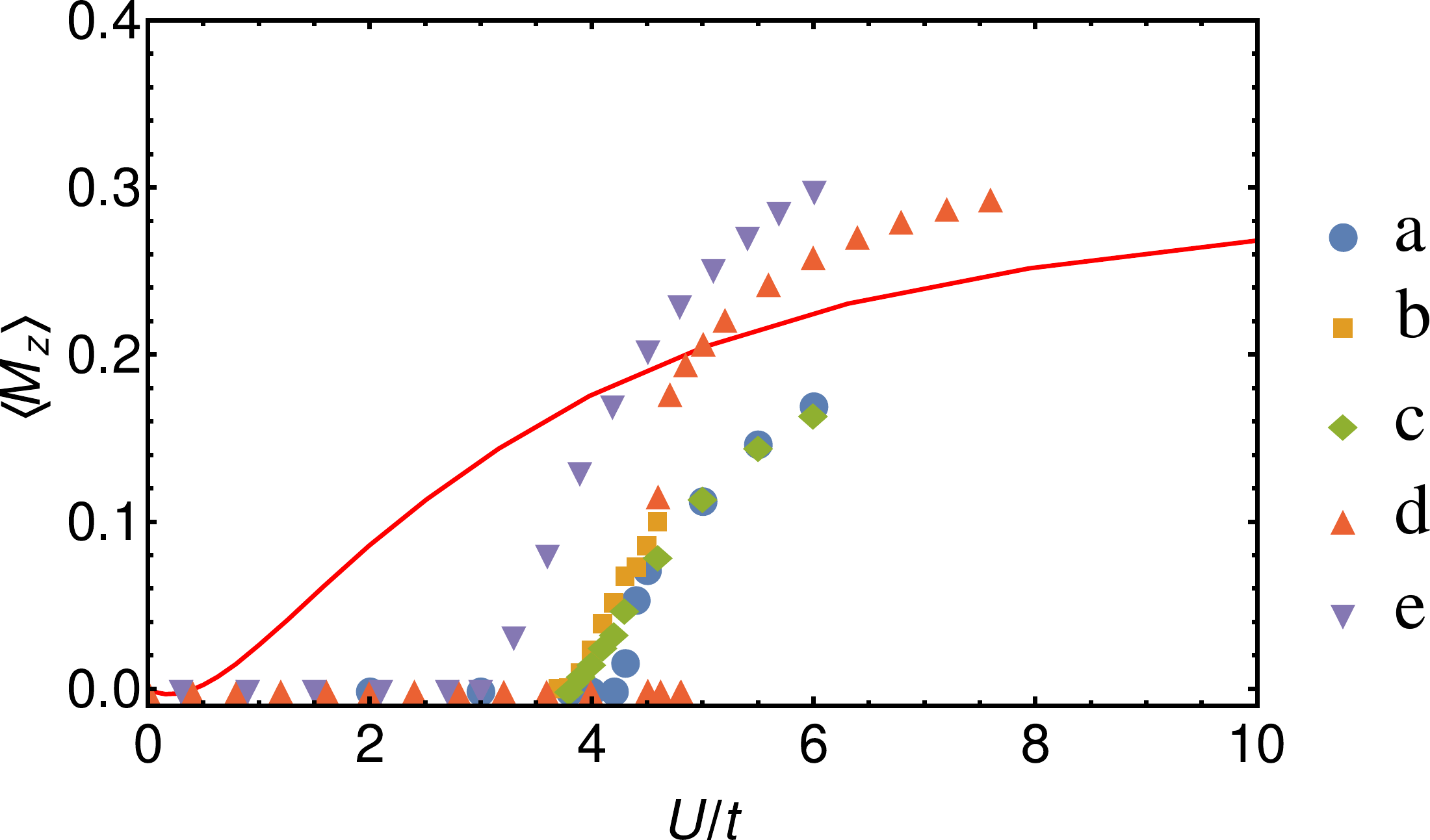}
\par\end{centering}

\caption{Comparison of our results (dashed and dotted line) 
to results presented in the literature: circles (a) Ref.~\cite{meng2010quantum},
squares (b) Ref.~\cite{sorella2012absence}, lozenges (c)
Ref.~\cite{assaad2013pinning}
(for a small magnetic field, the case $h_{0}=1$), 
triangles up (d)  Ref.~\cite{he2012cluster},  and triangles down (e) from
Ref.~\cite{chen2014intermediate}. All results are scaled so that complete sub-lattice magnetisation corresponds to a value of $\langle M_z \rangle=1/2$.
(See Fig.~\ref{fig:MZhx} for details of
our work).\label{fig:cfMzhx}}
\end{figure}

When we compare our results for the most sensitive parameter, the
sub-lattice magnetisation, to some recent results in the literature,
see Fig.~\ref{fig:cfMzhx}, we note first of all the similarity between
the literature results. The results from
Ref.~\cite{chen2014intermediate} are still subject to substantial
finite size corrections; and the results from
Ref.~\cite{chen2014intermediate} agree on the transition point, but not
on the nature of the transition and the size of the magnetisation
above the transition point. In the area where we can rely on our
results, which we estimate to be $U/t\gtrsim 6-8$, we find values of the
magnetisation entirely consistent with the literature.

\subsection{Excited states\label{sub:Results:Excited-states}}

\begin{figure}
\subfloat[square lattice.]{\includegraphics[width=6cm]{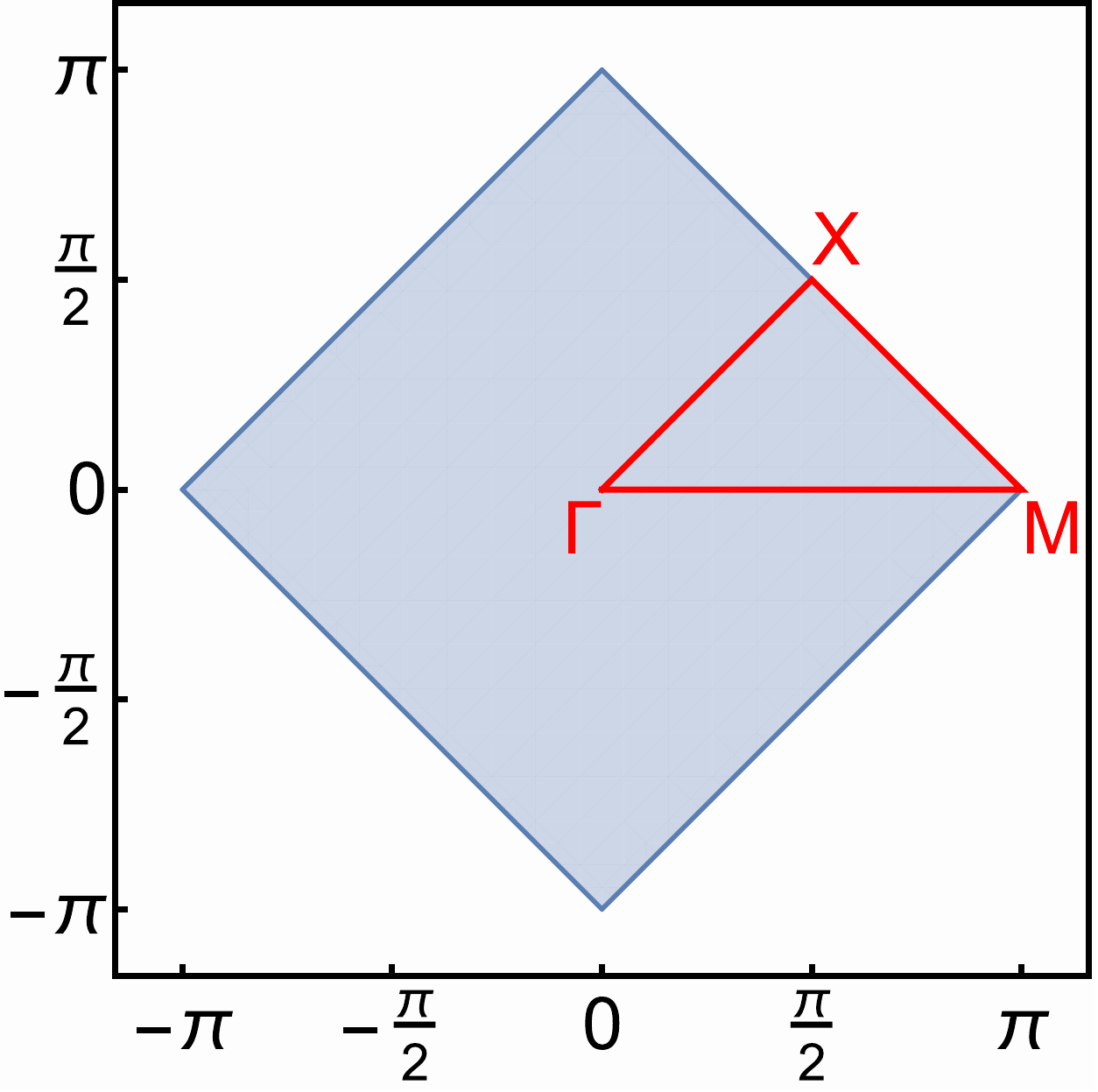}

}\subfloat[honeycomb lattice.]{\includegraphics[width=6cm]{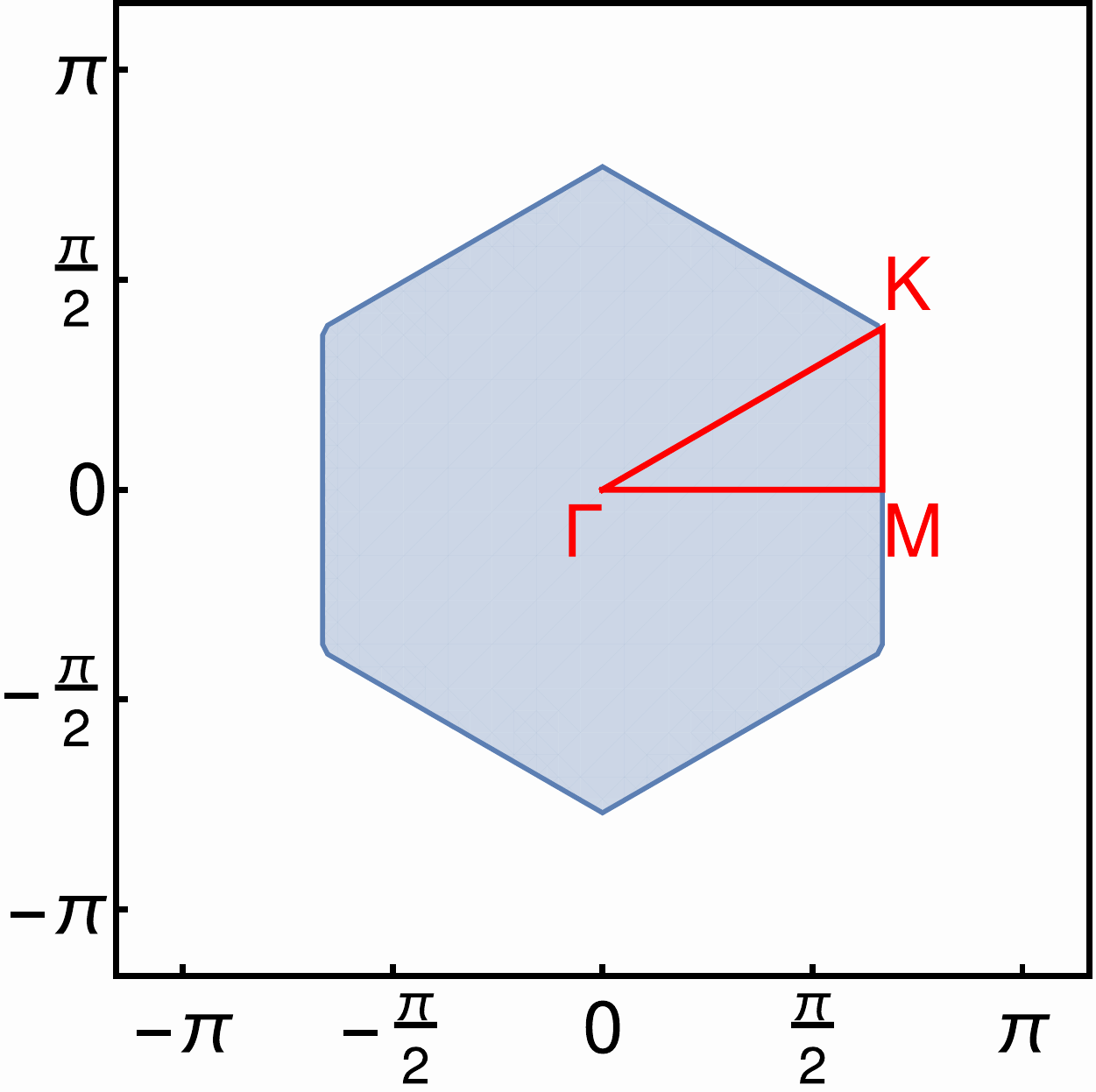}

}

\caption{The paths within the first Brillouin zone used in the presentation
of excitation energies in the remainder of this paper.\label{fig:Paths}}
\end{figure}

We now apply the method for excited states discussed in Secs. \ref{sub:Charge-gap}
and \ref{sub:Spin-excitations} to the Hubbard model. We label the
high symmetry points in the first Brillouin zone as in Figs.~\ref{fig:Paths}.

\subsubsection{Charge excitations}

\begin{figure}
\begin{centering}
\subfloat[Square lattice.]{\begin{centering}
\includegraphics[width=7cm]{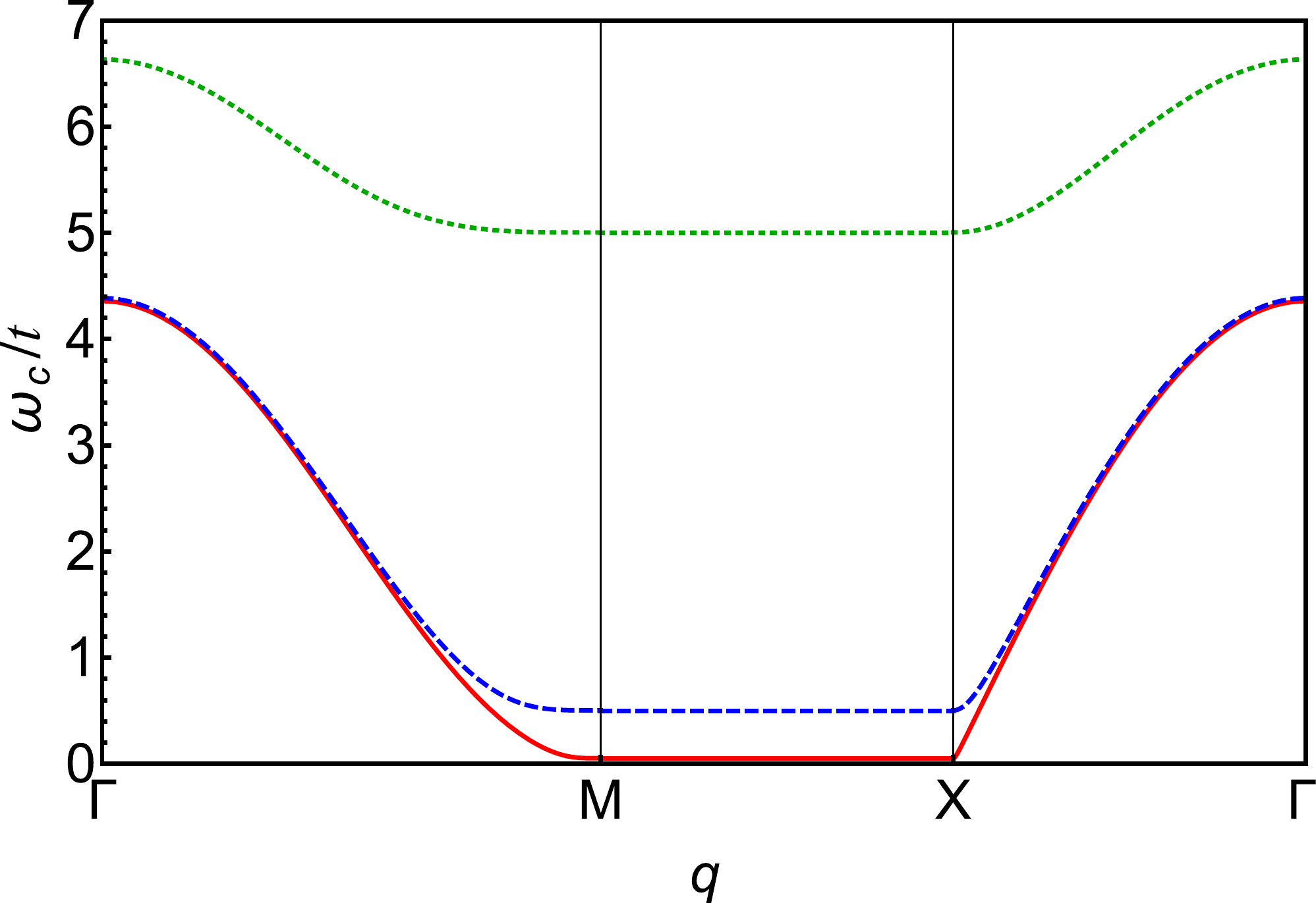}
\end{centering}}
\subfloat[Honeycomb lattice.]{\begin{centering}
\includegraphics[width=7cm]{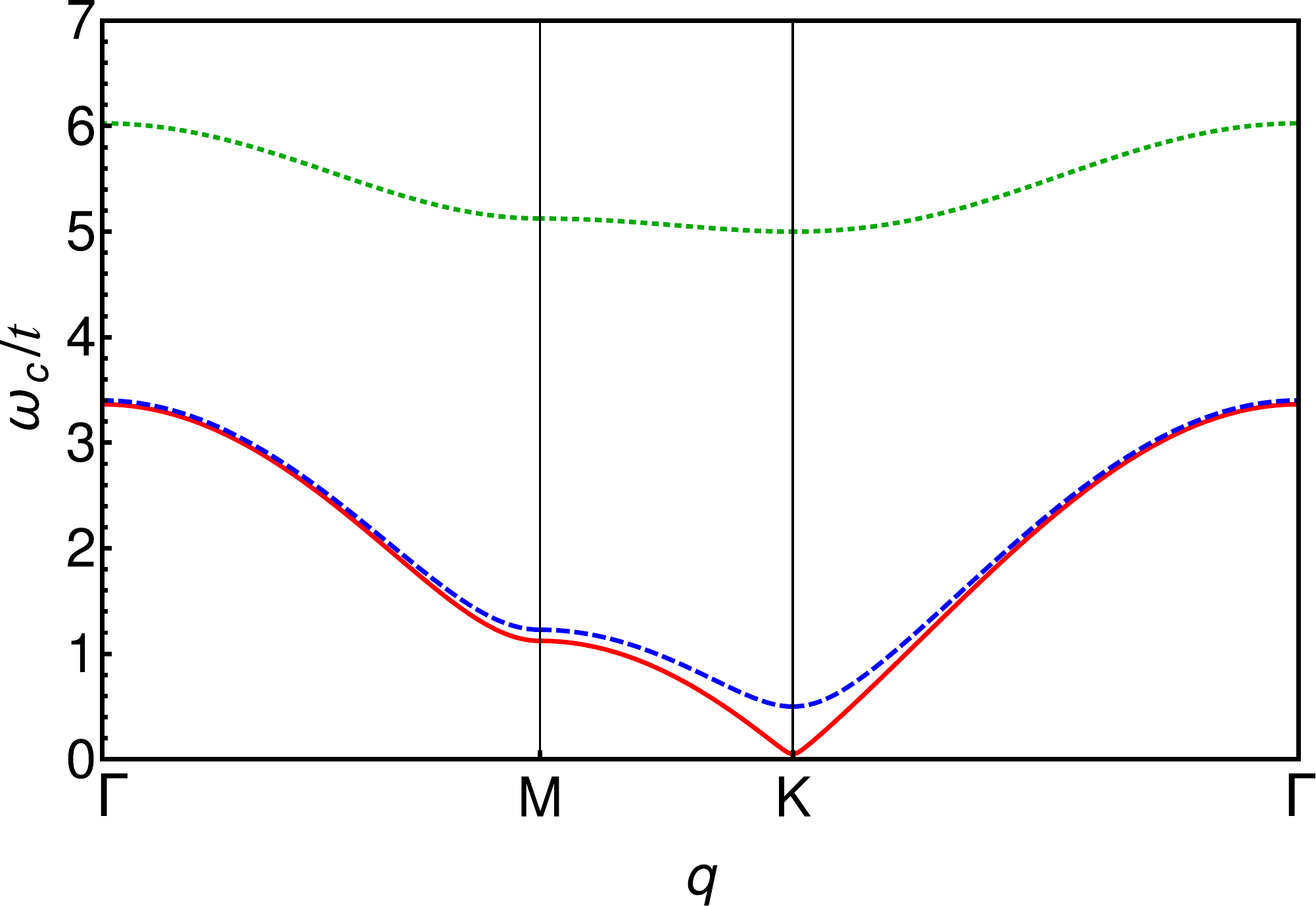}
\end{centering}
}
\end{centering}
\caption{Charge excitations in the 2D Hubbard model for $U/t=0.1$ (solid
red), $U/t=1$ (dashed blue) and $U/t=10$ (dotted green) for calculations
in the super-SUB1 approximation. \label{fig:Charge-gap}}
\end{figure}

The charge excitations, within the approximation made here, are 
rather similar to earlier results \cite{roger1990thecoupledcluster,
  bishop1995amicroscopic}, and thus also to those obtained with the mean-field method.
  The spectra show a large energy gap at large values of $U/t$: Since
  such excitations are suppressed in that limit, they scale as $U/t$.
 In Fig.~\ref{fig:Charge-gap} one can see an example of the
results. In the square lattice the frequency is constant along the
boundary of the first Brillouin zone (here represented as the line
$M-X$) with a value of $U/2$. In the honeycomb lattice we see the
dip around the $K$ point, which disappears as $U/t$ increases.

\subsubsection{Spin-flip excitations\label{sub:Results:Spin-gap}}

In the 1D case the spin-flip spectra can be calculated 
by solving the relevant eigenvalue problem on a lattice in $q$ space.
These show a striking similarity to the pictures for two-magnon excitations
in gapped 1D antiferromagnetic systems developed by Barnes \cite{barnes2003boundaries}.
This is due to the great similarity in the mathematical structure
of the problems, but the physics is very different! Our results are
for what is essentially a non-local \emph{single} magnon excitation
in the Hubbard model. The bound state at the bottom of the spectrum
corresponds to the local
single magnon excitation in the Hubbard model (in the large $U/t$ limit).
As we can see the description of the magnon mode is not completely
satisfactory; even though the continuum moves far away in this limit,
the single bound state has a constant energy $\frac{t^{2}}{U}2z(1+\alpha_{1}^{\Delta})$,
as specified by Eq.~(\ref{eq:ExLargeU}). On rather general grounds
we do expect a gapless magnon in this limit \cite{peres2004phasediagram};
without the Heisenberg corrections
our result is only equal to the amplitude of the magnon spectrum
for the Heisenberg model.

\begin{figure}
\begin{center}
\subfloat[$U/t=2.5$]{\includegraphics[width=6cm]{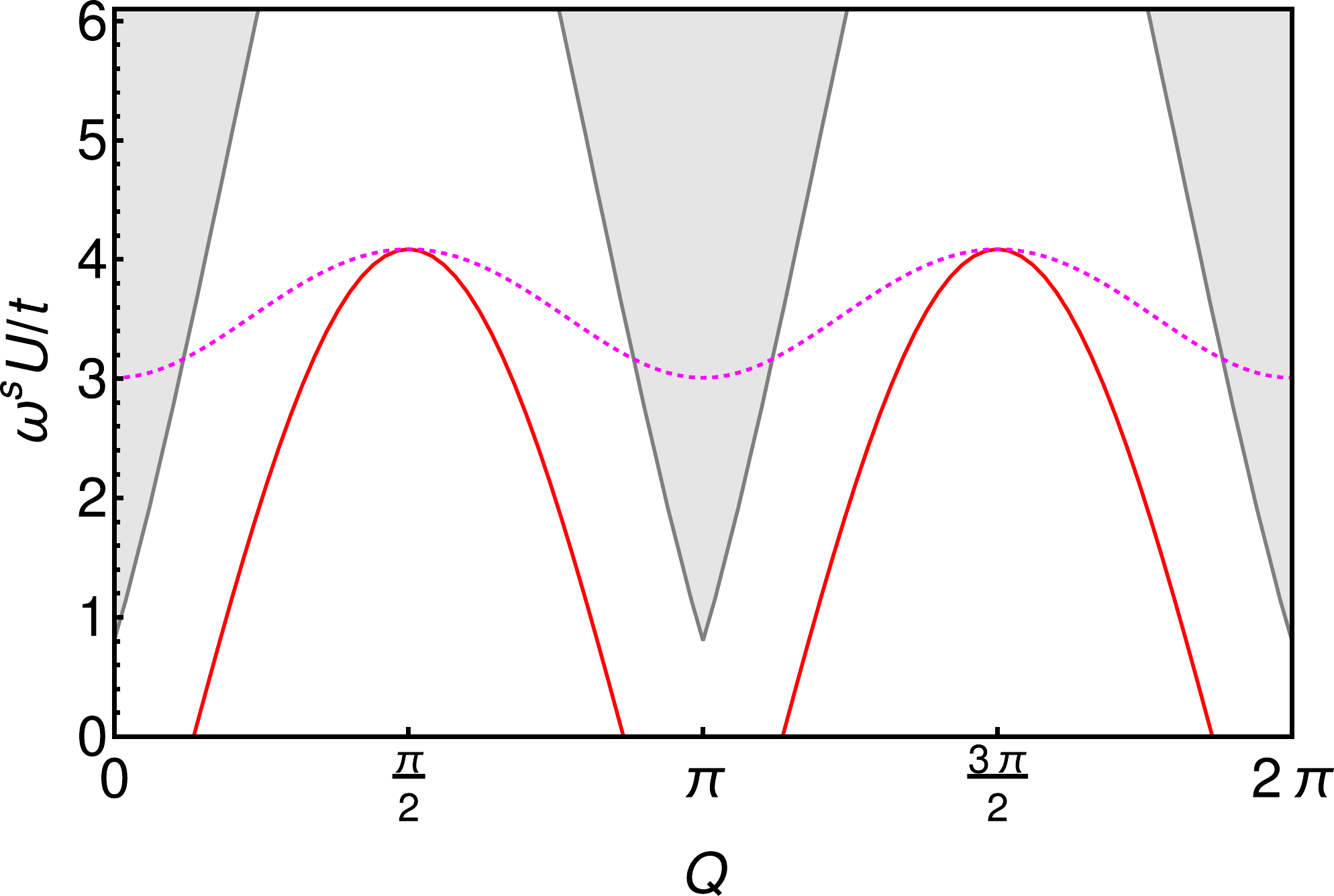}}
\subfloat[$U/t=5$]{\includegraphics[width=6cm]{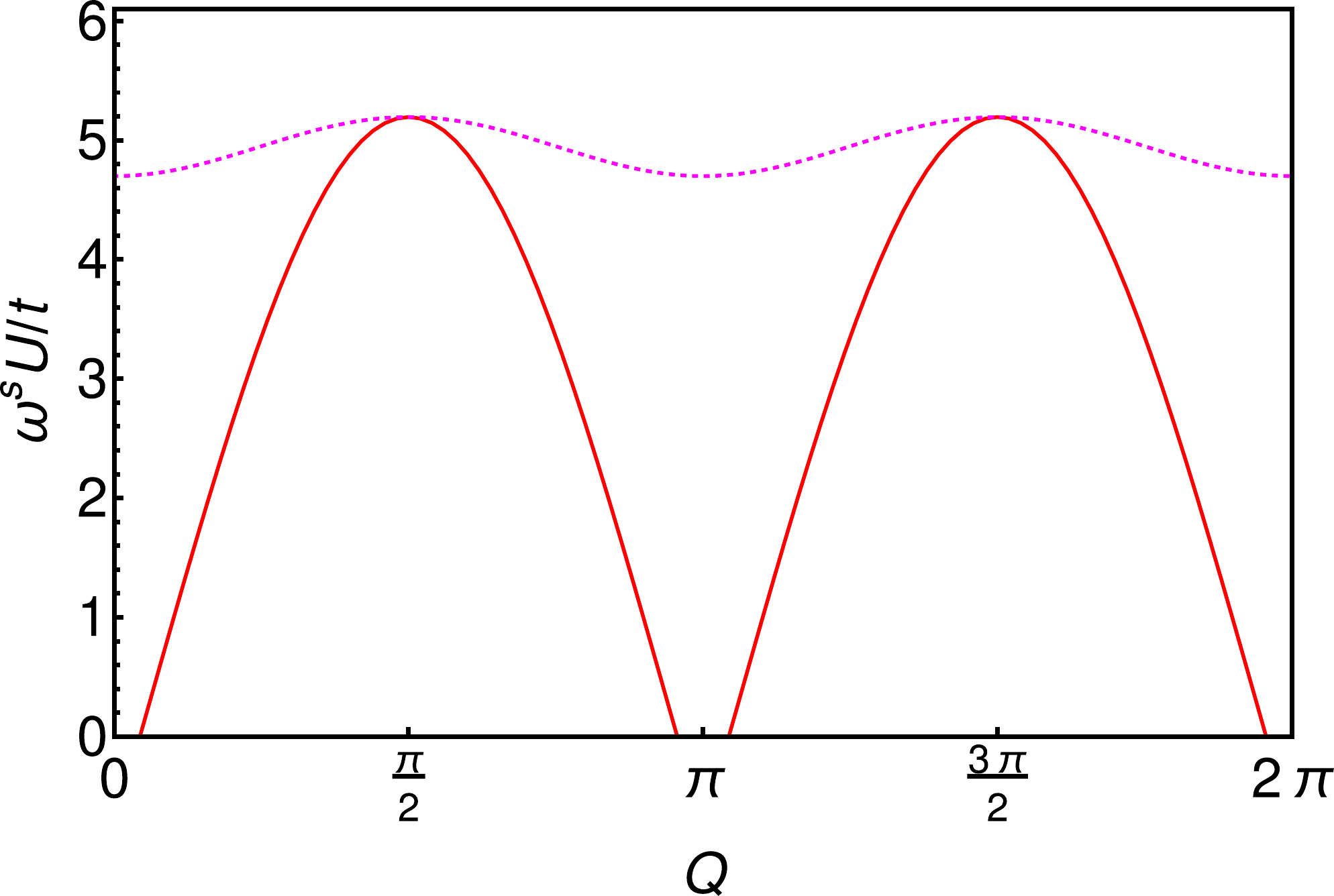}}\\
\subfloat[$U/t=10$]{\includegraphics[width=6cm]{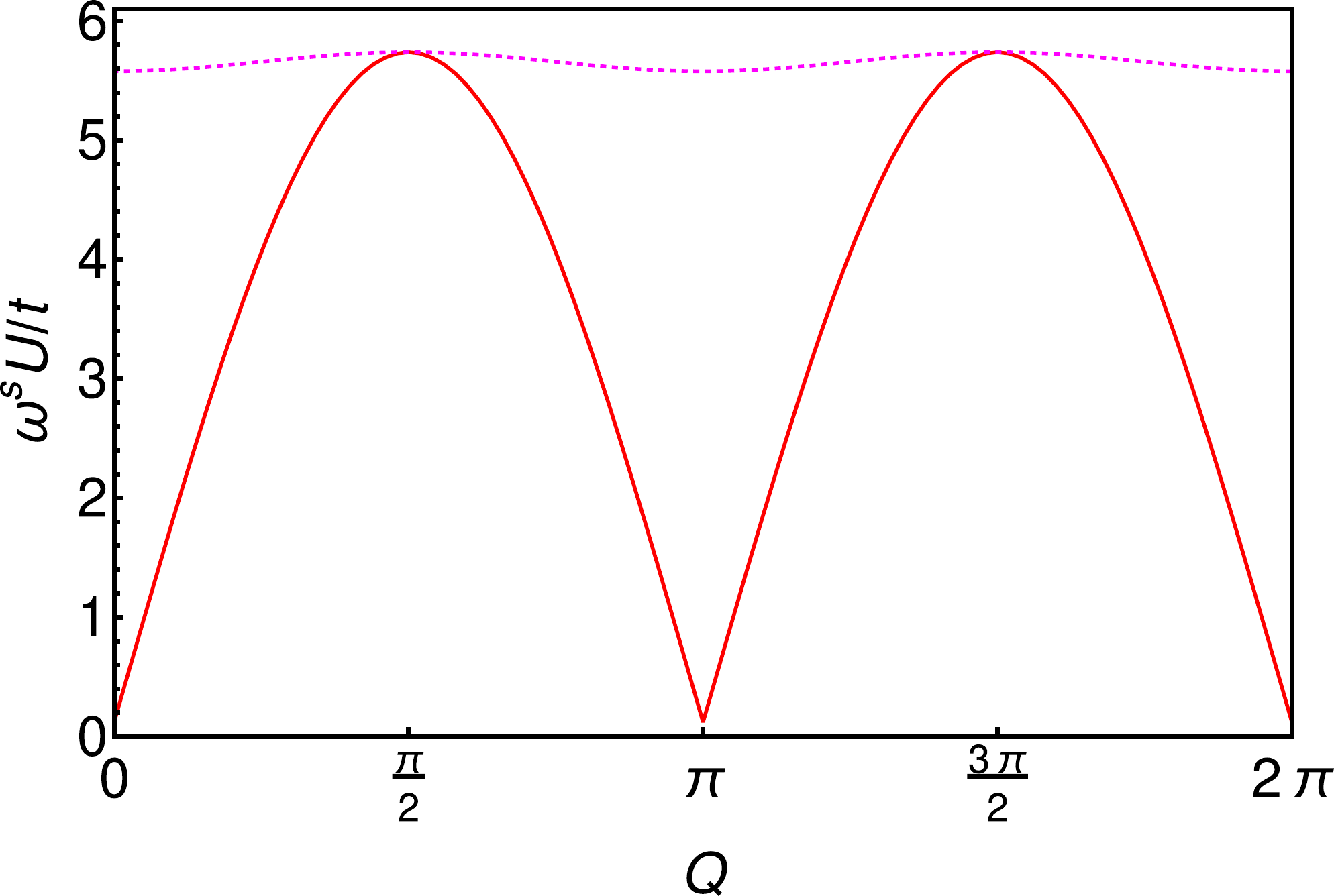}}
\subfloat[$U/t=20$]{\includegraphics[width=6cm]{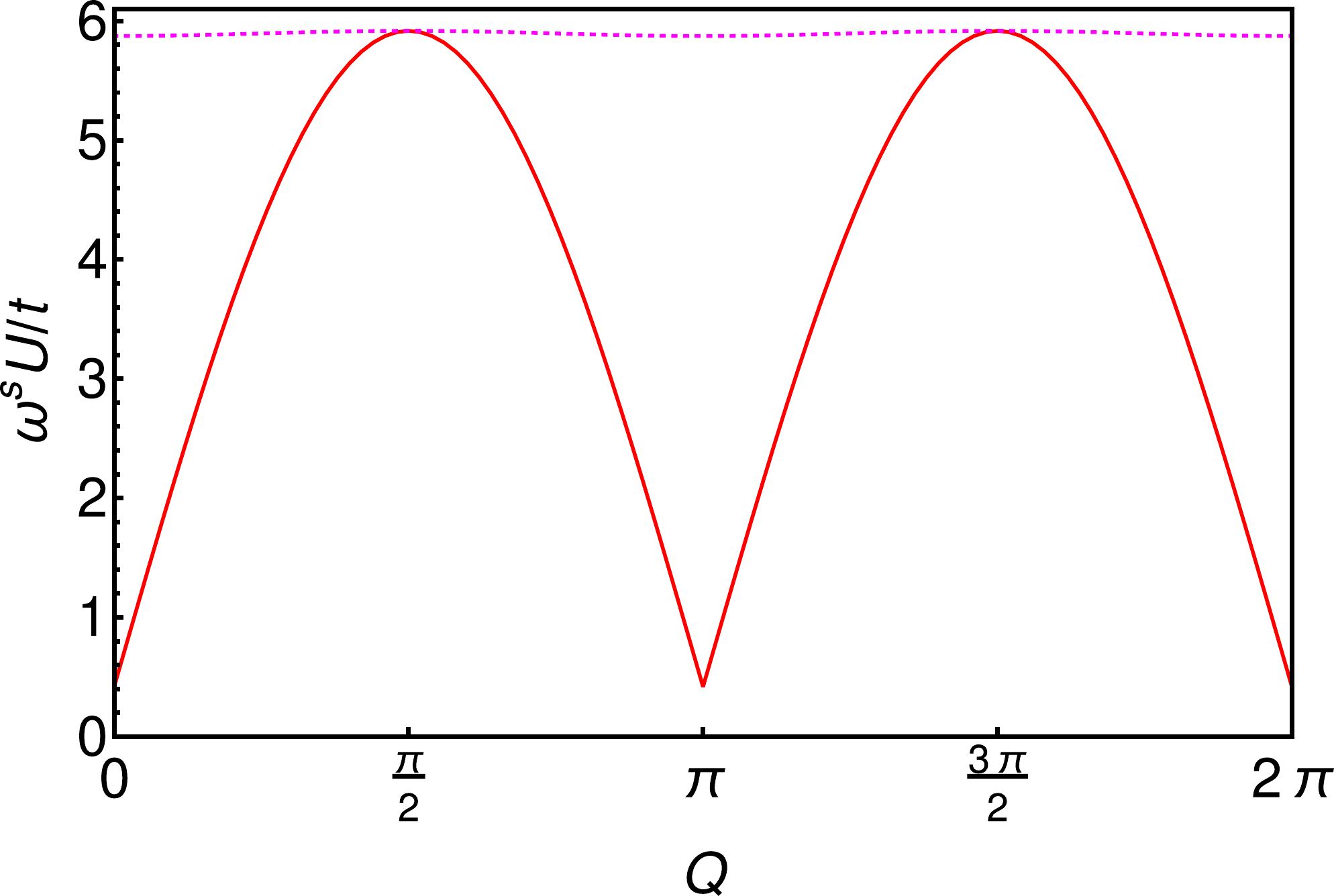}}\\
\subfloat[$U/t=100$]{\includegraphics[width=6cm]{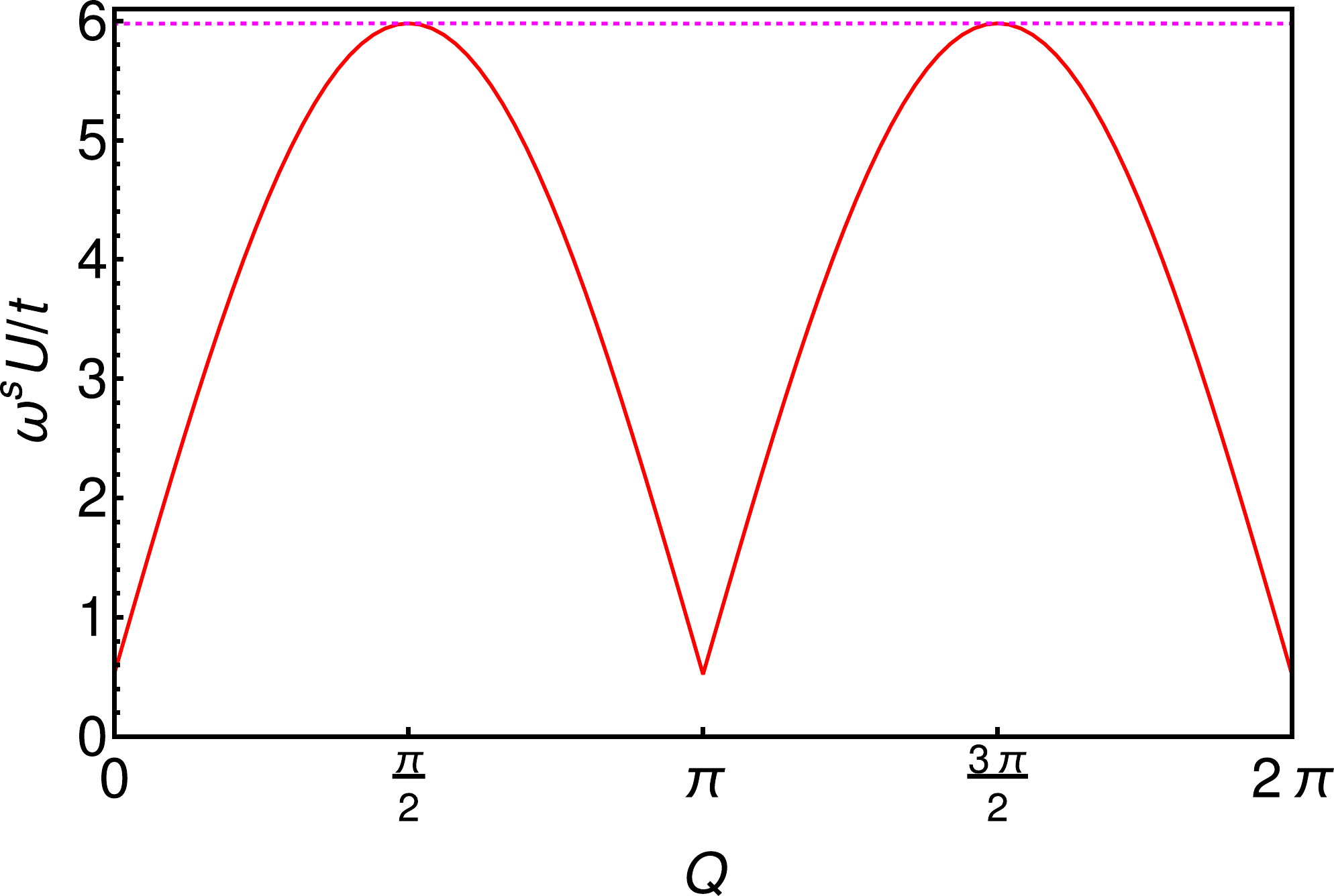}}
\end{center}

\caption{The spin-excitation spectrum for $S_{z}=1$ in the 1D Hubbard model
for several values of $U/t$. In each case we show the uncorrected approximation
(dotted magenta line) and the Heisenberg corrected results for $\Delta=\Delta_c$
(red solid line). The grey area is the continuum of states visible for small $U/t$.
\label{fig:Ex1Da}}
\end{figure}

We have already discussed how we can correct for some of these
shortcomings when we analyse the Heisenberg model; if we add the
corrections discussed in Sec.~\ref{sec:LLHeis} we should get much
better answers for small $t/U$.  As we can see in Fig.~\ref{fig:Ex1Da},
this is indeed the case. These corrections give a sensible magnon
spectrum up to $U/t\simeq 5-10$, after which things break
down. Interestingly, this is also roughly the range of parameters
where the separation between continuum and bound state becomes
comparable to the bound state energy.






Clearly in applying the super-SUB1 approximation, which is designed to
improve results at large $U/t$, we pay a price at smaller values of
$U/t$. As can be seen in Fig.\ \ref{fig:Hub1d}a, we find a slight overbinding
for large $U/t$, but this becomes a large effect for small $U/t$. The
exact solution is bracketed between the $\Delta=1$ and the
$\Delta=\Delta_c$ results down to $U/t\approx 2$, suggesting that the
approximation we make gives a substantial improvement when we take
$U/t$ above that value.

\begin{figure}
\begin{center}
\subfloat[$U/t=3$]{\includegraphics[width=6cm]{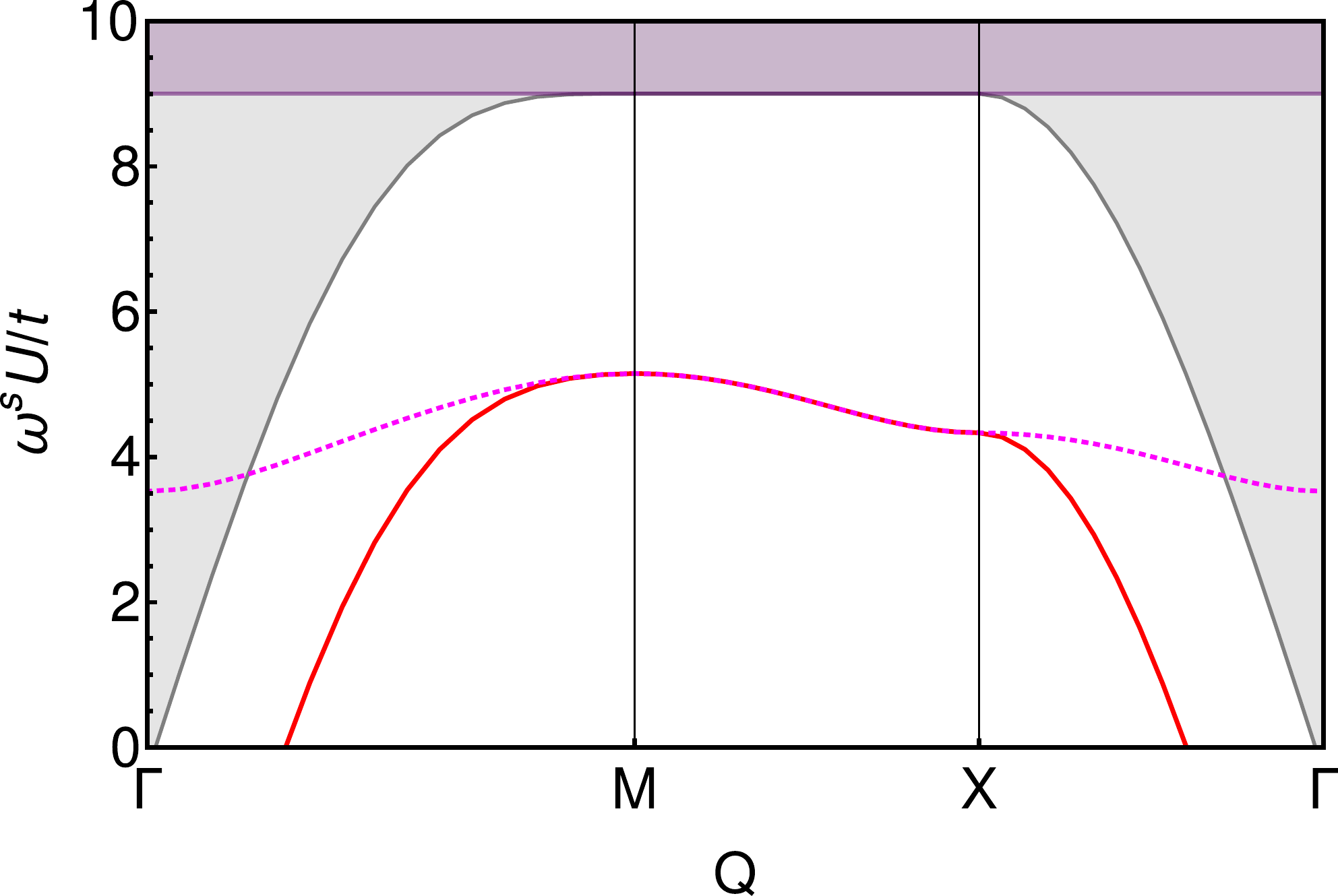}}
\subfloat[$U/t=5$]{\includegraphics[width=6cm]{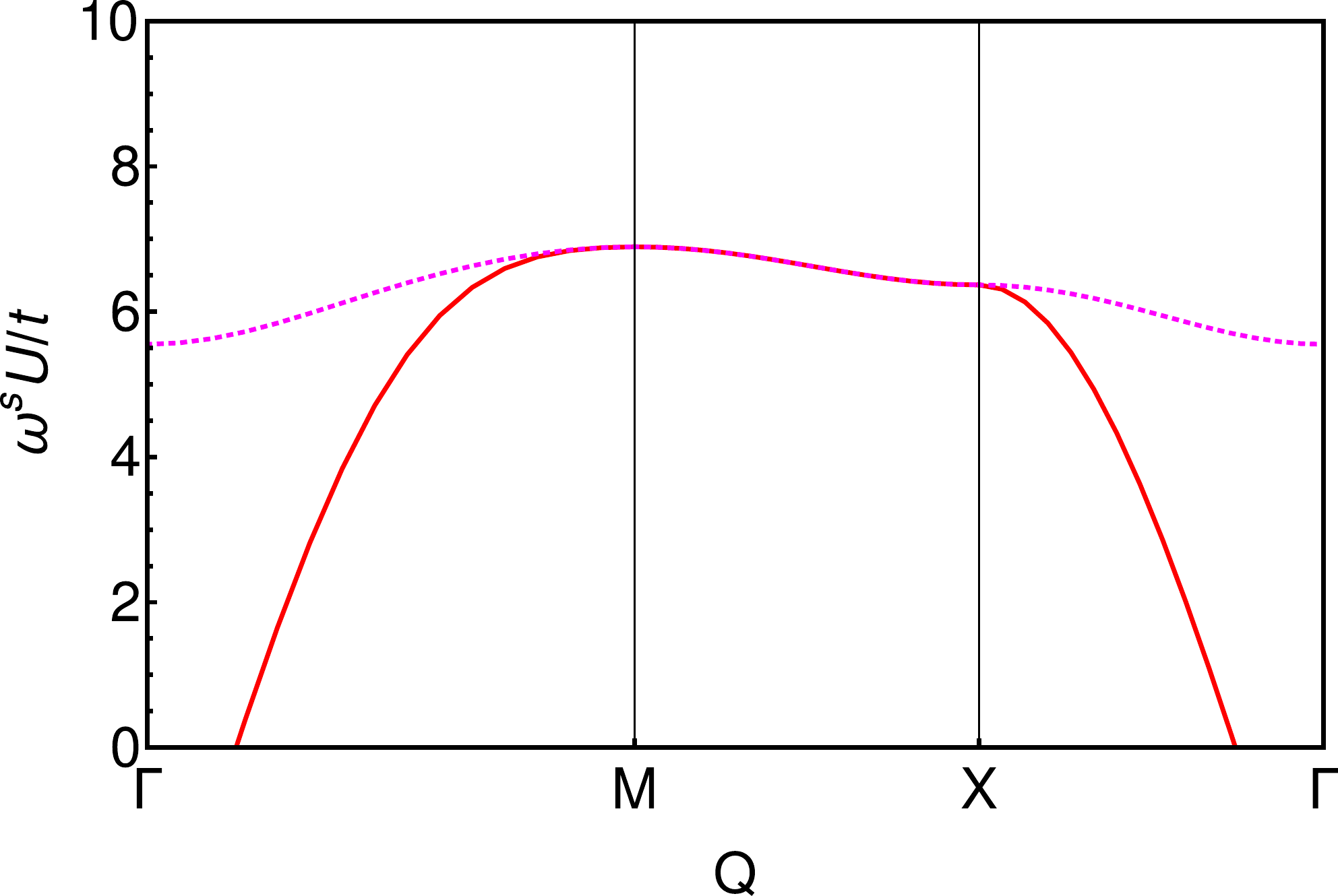}}\\
\subfloat[$U/t=10$]{\includegraphics[width=6cm]{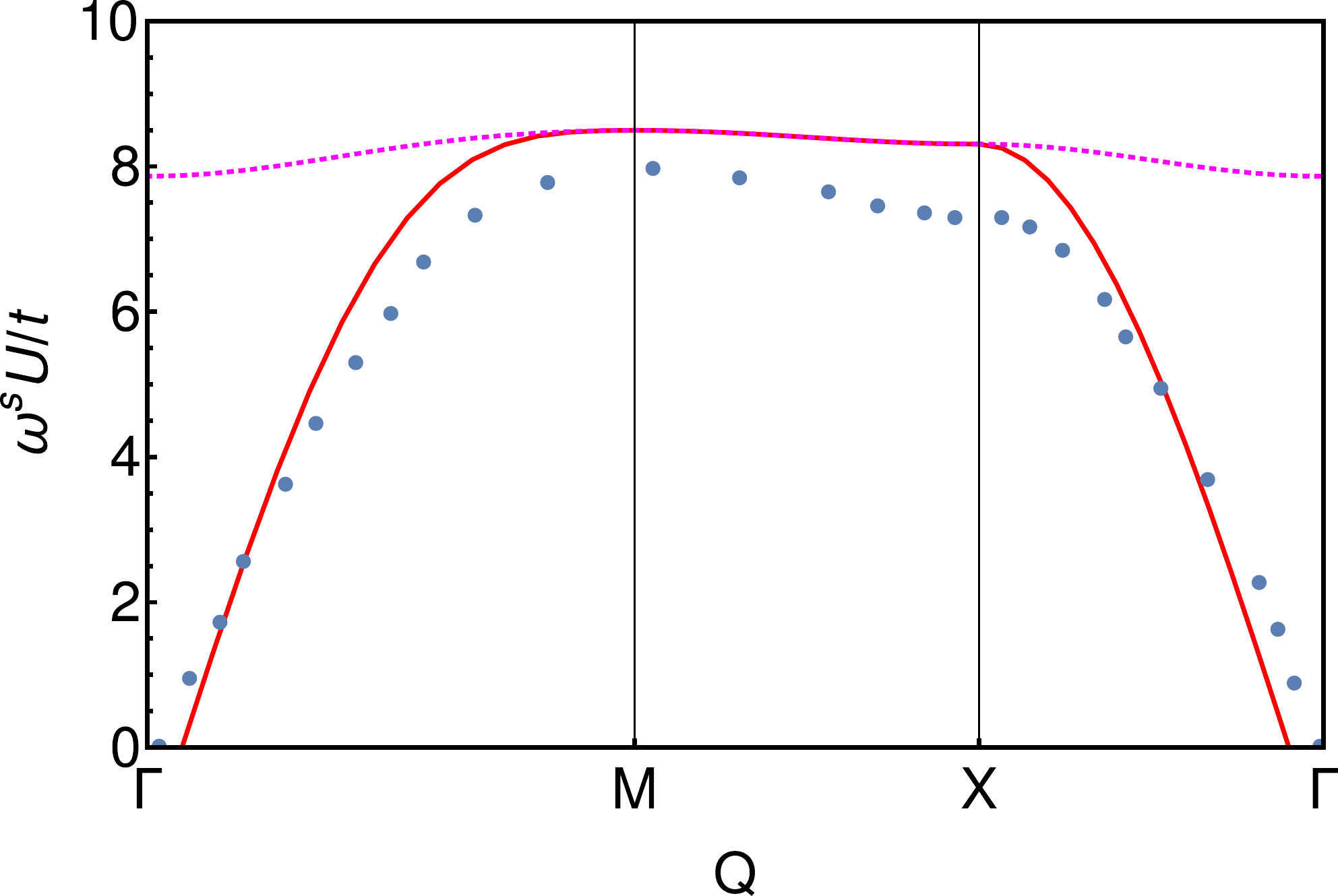}}
\subfloat[$U/t=20$]{\includegraphics[width=6cm]{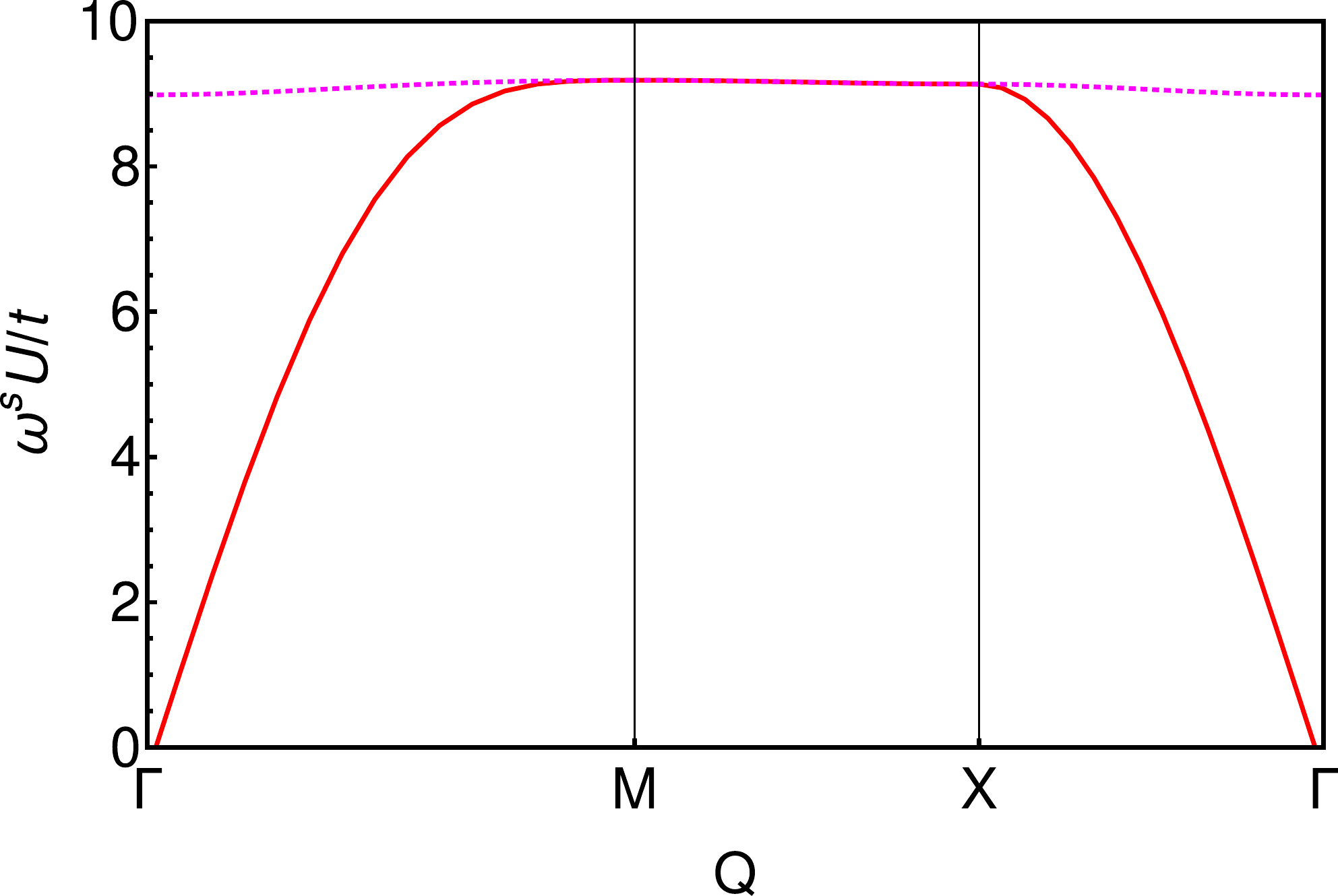}}\\
\subfloat[$U/t=100$]{\includegraphics[width=6cm]{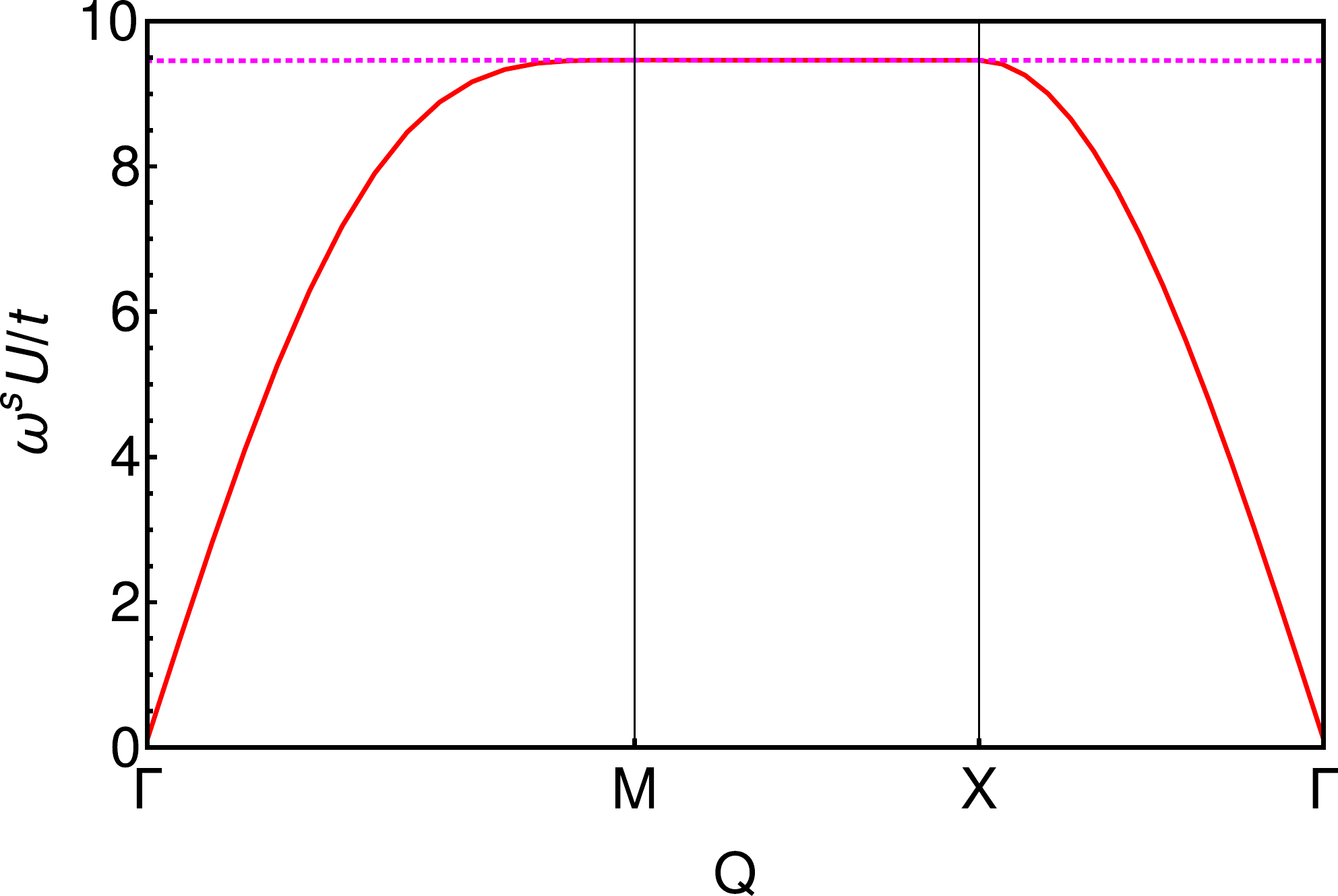}}
\end{center}
\caption{The spin-excitation spectrum for $S_{z}=1$ for the 2D Hubbard model
on the square lattice
for several values of $U/t$. In each case we show the uncorrected approximation
(dotted magenta line) and the Heisenberg corrected results for $\Delta=\Delta_c$
(red solid line). The grey area is the continuum of states visible for small $U/t$.
The data points for $U/t=10$ are from the series expansion of Ref.~\cite{zheng_magnon_2005} for $U/t=10.5$. We have suppressed the error bars on the series expansion results.
\label{fig:2dsquareex}}
\end{figure}

Since the structure of the approximation schemes is so similar, even though  the
results  for the magnetisation looks rather different, it seems 
reasonable to assume that we obtain reliable
results for the 2D problems for similar values of $U/t$, maybe slightly large to be on
the conservative side.  We first
look at the square lattice, Fig.\ \ref{fig:2dsquareex}. As in the 1D
model we see a continuum, and a bound state that merges with the
continuum for small $U/t$. The continuum is flat at the lower end of
the spectrum--actually since this caused by the combination of states
at the edge of the Brillouin zone, the energy is exactly $U/t$. If we compare to
the series expansion results from Ref.\ \cite{zheng_magnon_2005}, who seem to
have taken a similar approach to incorporating the Heisenberg model, we
see that our results are very close to theirs--the difference is however
larger than
the quoted error-bars. 
Also, our spectrum is substantially flatter 
on the zone boundary between M and X; this can be traced to the flatness of 
the charge excitation spectrum in the CCM approximation: the series expansion 
has more structure on that boundary. It may well be that if we include higher order
operators in the charge-state caclculations this result would improve substantially.

\begin{figure}
\begin{center}
\subfloat[$U/t=3$]{\includegraphics[width=6cm]{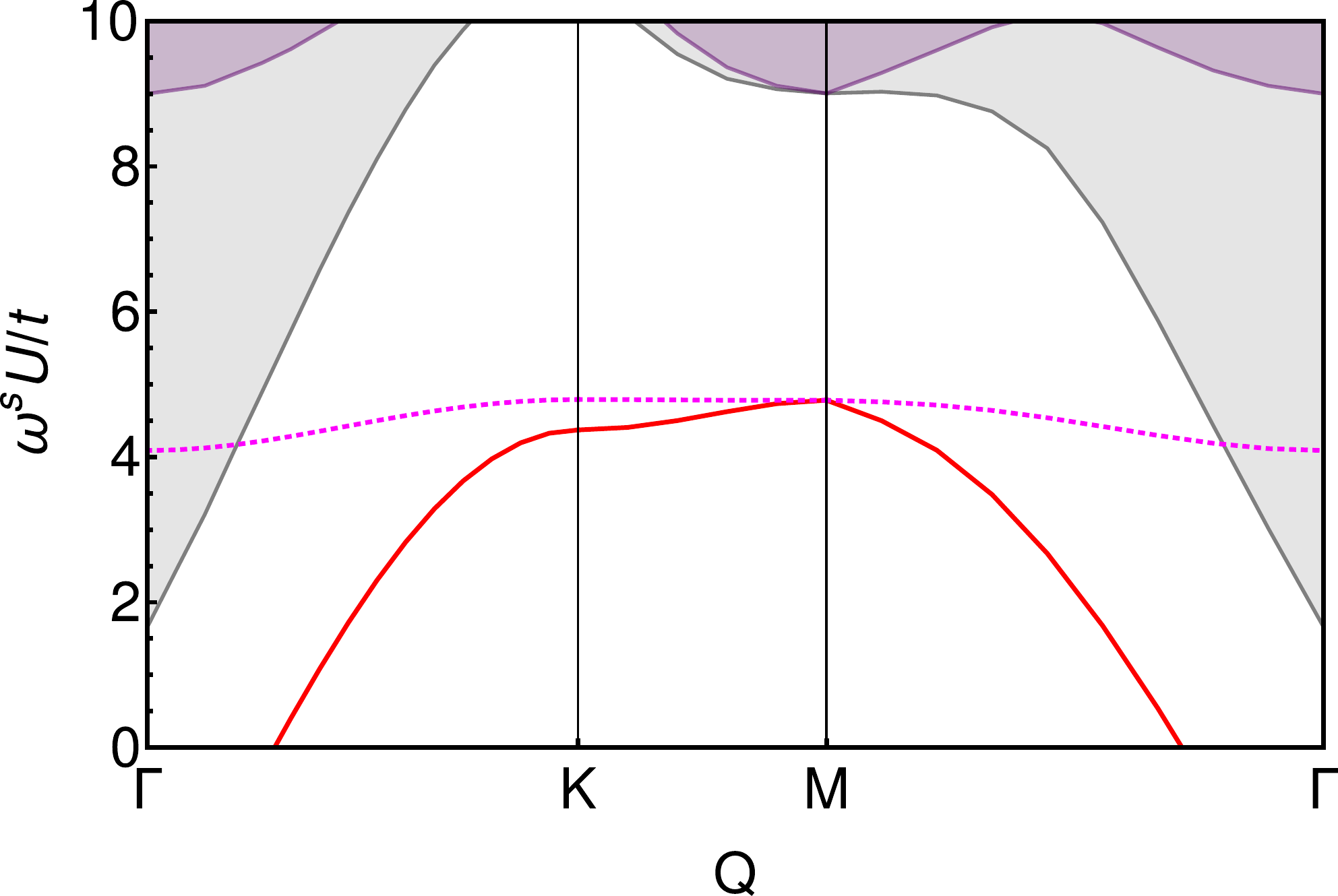}}
\subfloat[$U/t=5$]{\includegraphics[width=6cm]{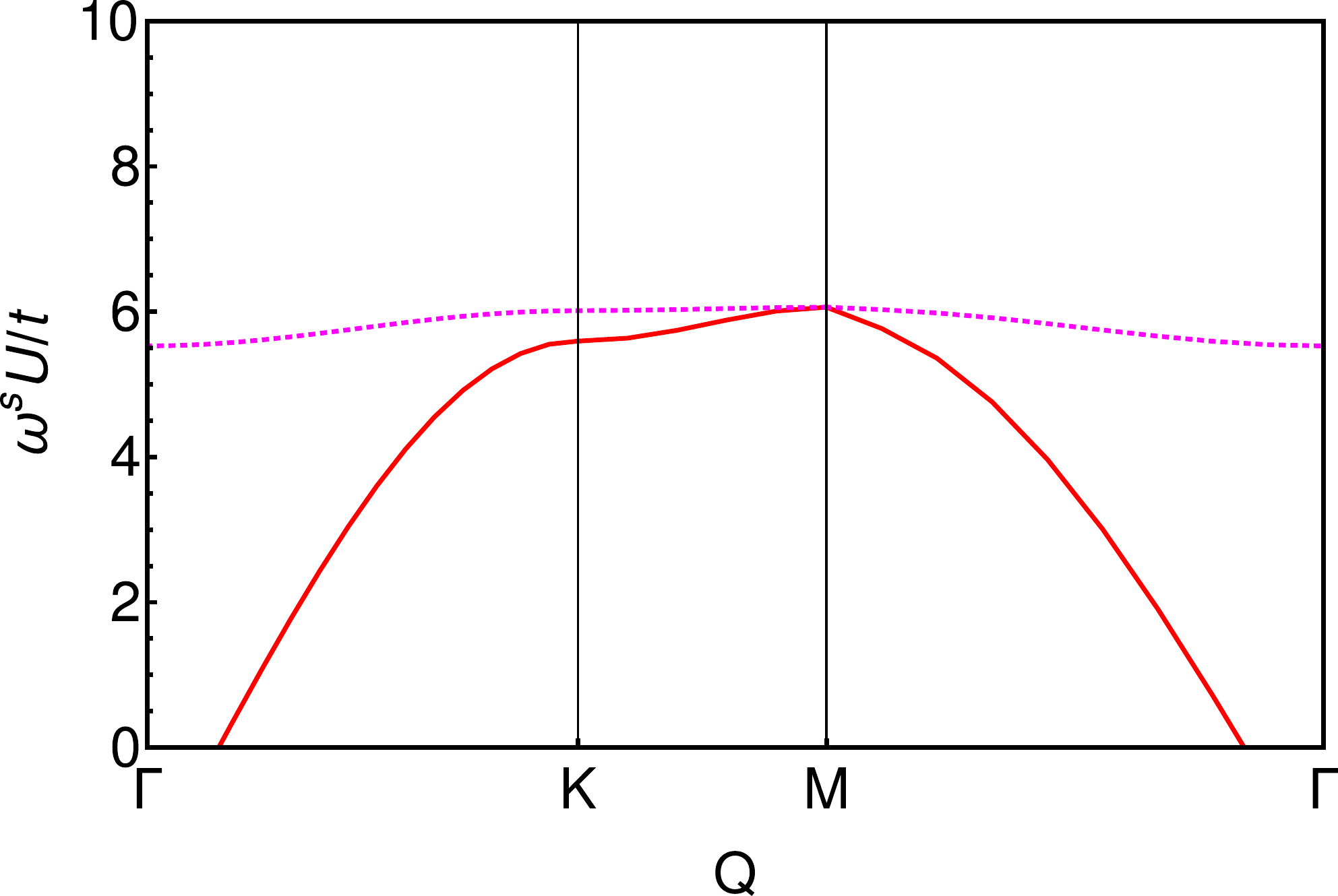}}\\
\subfloat[$U/t=10$]{\includegraphics[width=6cm]{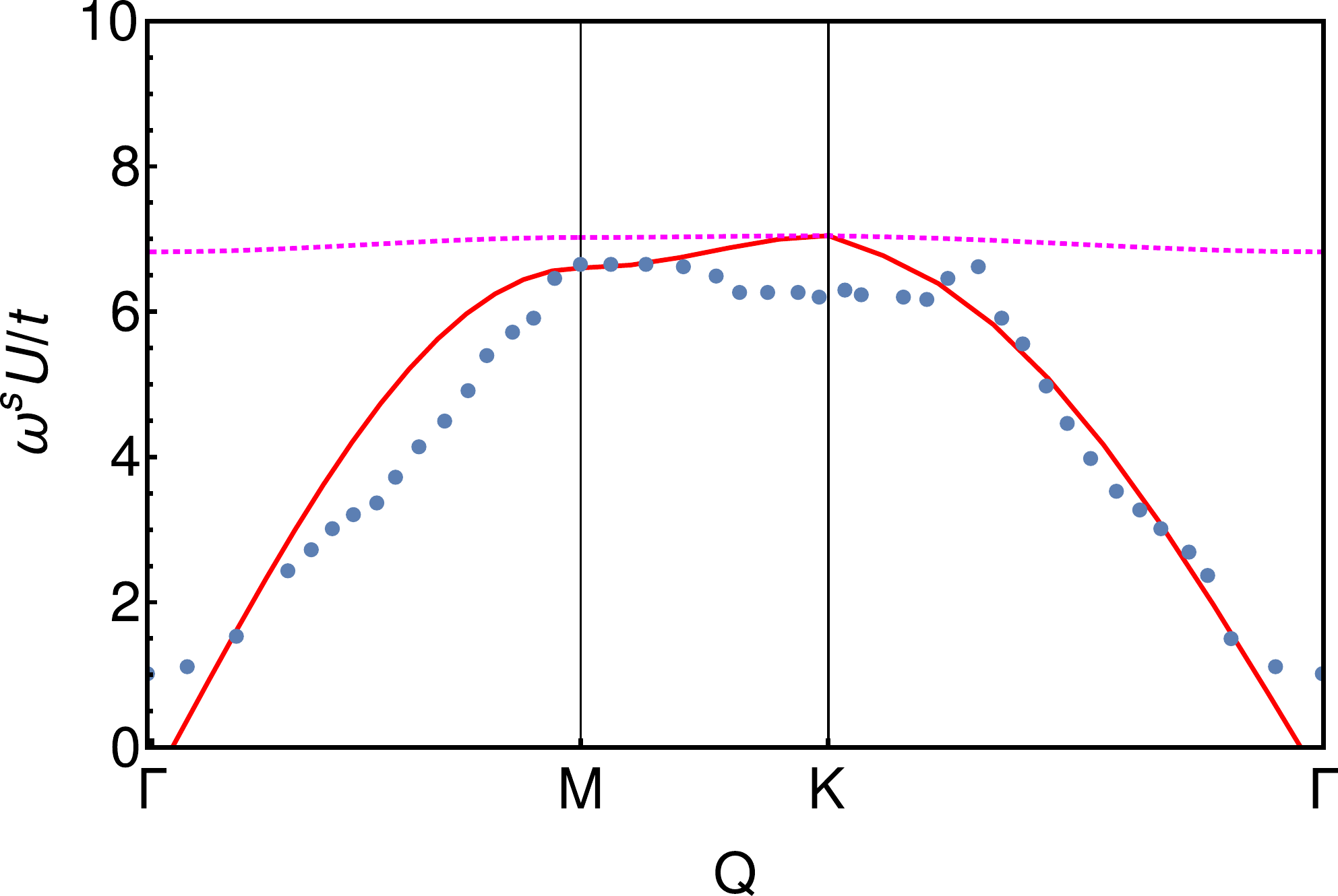}}
\subfloat[$U/t=20$]{\includegraphics[width=6cm]{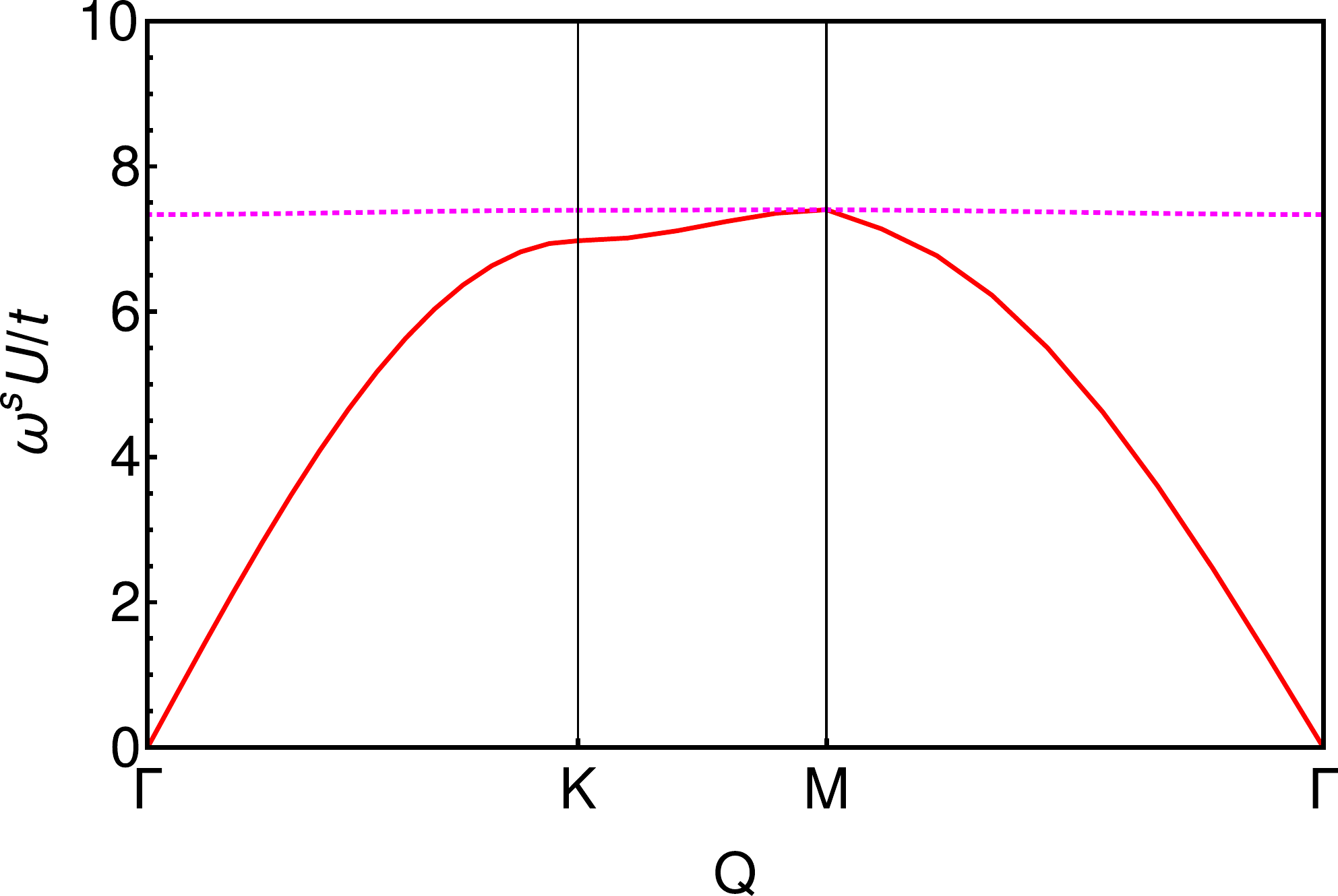}}\\
\subfloat[$U/t=100$]{\includegraphics[width=6cm]{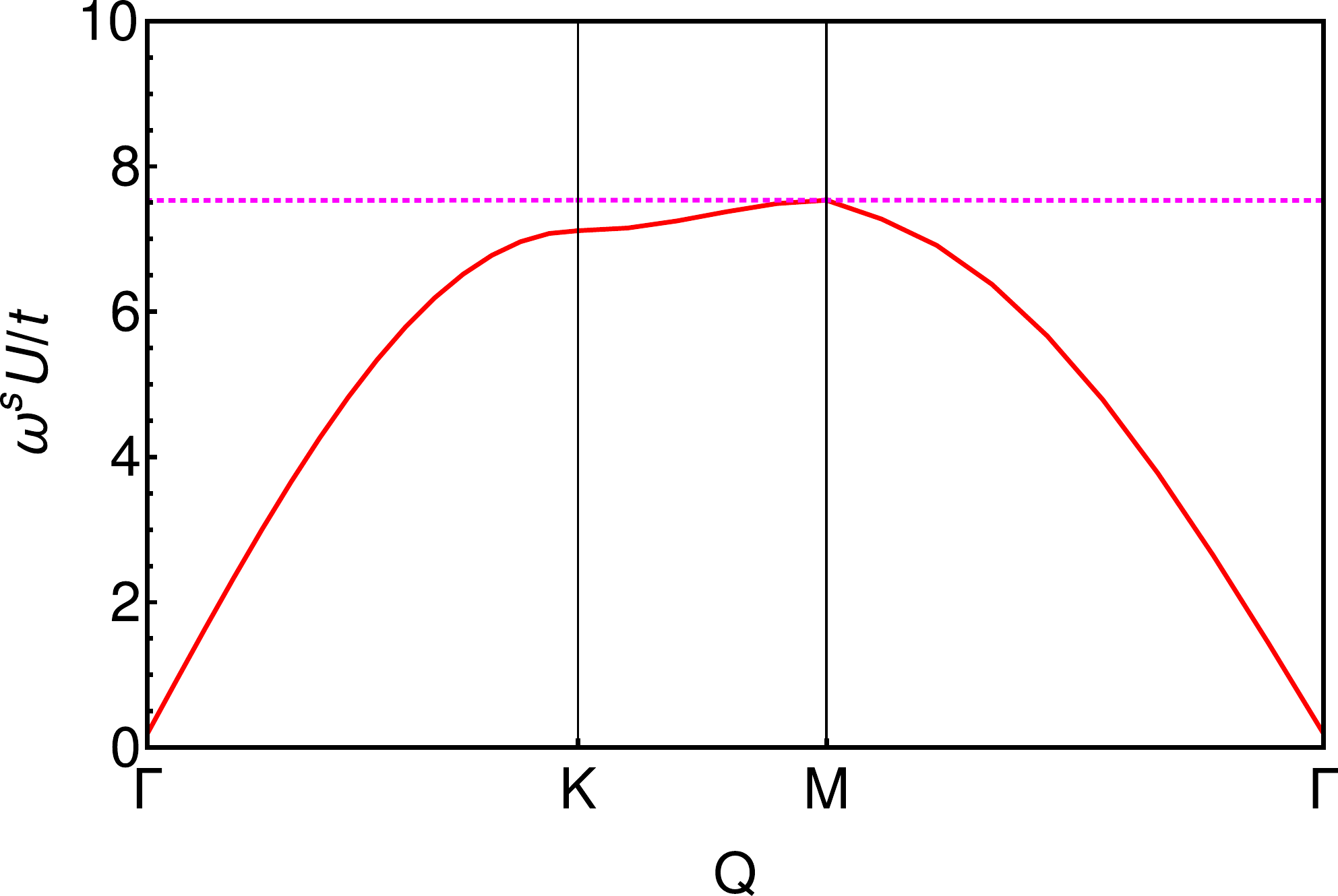}}
\end{center}
\caption{The spin-excitation spectrum for $S_{z}=1$ for the 2D Hubbard
  model on the honeycomb lattice for several values of $U/t$. In each
  case we show the uncorrected approximation (dotted magenta line) and
  the Heisenberg corrected results for $\Delta=\Delta_c$ (red solid
  line). The grey area is the continuum of states visible for small $U/t$.
  The data points for
  $U/t=10$ are from the series expansion of
  Ref.~\cite{paiva_ground-state_2005} for the same value of $U/t$. We have suppressed
  the substantial error bars on the series expansion results.
\label{fig:2dhoneycombex}}
\end{figure}

If we do the same thing for the model on the honeycomb lattice,
Fig.\ \ref{fig:2dhoneycombex}, we see a slightly different
behaviour. There still is a continuum and a bound state, but the
continuum band now has  structure. Once again, we see that the
bound state converges to zero for large $U/t$.
 If we compare to
the series expansion results from Ref.\ \cite{paiva_ground-state_2005}, we
see that our results are very close to theirs--the error bars on the series expansion are substantial. 
There may be a bit more structure in the series expansion results, but we are not convinced
all of this structure is real--it is not mirrored in the behaviour of the charge excitations.

\section{Outlook and Conclusions\label{sec:Outlook-and-Conclusions}}

In this paper we have investigated the ground and excitation state
properties of Hubbard models in one and two-dimensions using the CCM,
in similar fashion as earlier CCM analysis for the spin models.
As expected, the analysis for the Hubbard models is much more involved
than those of the spin models due to inclusion of the charge
fluctuations, in addition to the spin fluctuations. Even though there
is a close parallel between Hubbard and Heisenberg models
for large $U/t$, we have concluded that
the SUB2 scheme for the ground states of the spin models corresponds
to the SUB3 scheme in the Hubbard model. Similar conclusion can also
be drawn for the CCM analysis for the spin-flip excitations.

For efficiency purpose, we have directly employed the results of the
two-body spin-spin correlations from our earlier calculations of the
spin $XXZ$ model with the critical anisotropy in our evaluation of
the ground state of the Hubbard models, avoiding explicit analysis
of the complex SUB3 scheme, and have obtained reasonably good numerical
results for the ground-state energies, the sub-lattice magnetization,
and the charge excitation spectra for wide range values of the on-site
interaction parameter $U/t$, when comparing with the corresponding
results by numerical Monte Carlo methods. In the large-$U/t$ limit,
our results reduce to those of the spin models as expected.

For the spin-flip excitation states, however, we do not obtain the
corresponding gapless spin-wave spectrum as we would expect in the
large-$U/t$ limit. Instead, we have obtained gapped spectra which becomes
flat in the large-$U/t$ limit. We have concluded that this problem
of gapped spectra can be solved by inclusion of the higher-order correlations
in the excitation operators in a similar way as
for the ground state, again due to the presence of charge fluctuations
in the Hubbard models. More specifically, we need to consider mode-mode
couplings as mentioned in Sec.~\ref{sub:Results:Spin-gap}. 
We have only dealt with this problem for large $U$;
nevertheless our results for the spin-flip excitation states show well-defined
bound states below a continuum for large values of $U/t$, with
an the amplitude equal to the corresponding spin-wave velocity. The approximation breaks down as the energy of this
bound state drops below the ground state near the $\Gamma$ point
as $U/t$ decreases. We expect that this effect disappears
when higher-order correlations in the excitation operator are
included and we hope to report results in the near future.

It is also interesting to apply the extended CCM (ECCM) analysis to
the Hubbard model since the lowest approximation of the ECCM will
reproduce the mean-field results, to investigate in particular the
properties of the metal-insulator transition for the honeycomb lattice
model. We are also planning to extend our analysis to alternative
models with many interesting phase structures. The Kitaev-Heisenberg
model \cite{kitaev2006anyonsin} has been generalised to a Hubbard-type
model in optical lattices \cite{duan2003controlling} and studied
in detail by Hassan and collaborators \cite{hassan2013stablealgebraic,faye2014topological}.
The reason for the interest is the potential for realising an algebraic
spin liquid. The only modification we need to make to to the Hubbard
model is to make the hopping term spin dependent, which could be,
in principle, implemented as a straightforward extension to the work
reported here. 
\begin{acknowledgments}
One of us (WAA) would like to acknowledge the Higher Committee for
Education Development in Iraq (HCED) for support through a scholarship.
\end{acknowledgments}
\appendix
\section{Detailed derivation of CCM equations\label{app:A}}
\subsection{SUB2 on-site approximation\label{App:A-SUB2OS}}
The one- and two-body CCM equations now become much simpler, and we find
\begin{eqnarray}
tz\left(\gamma_{\vec{q}}\left(1+s_{1}^{(1)}\right)-\gamma_{-\vec{q}}s_{\vec{q}}^{2}\right)+Us_{\vec{q}} & = & 0,
\end{eqnarray}
\begin{align}
-U2\left[\left(s_{\vec{j}_{1}-\vec{i}_{1}}\right)^{2}\right]-4t\left(\sum_{\vec{\rho}}s_{\vec{\rho}}\right)s_{\vec{j}_{1}-\vec{i}_{1}}^{(1)} & =0\implies\nonumber \\
s_{\vec{r}}^{(1)} & =-\frac{1}{k}\frac{1}{s_{1}}s_{\vec{r}}^{2},
\end{align}
where we solve for $s^{(1)}$. 

We can also derive a similar set of equations for the coefficients in
$\tilde{S}$ which are needed to evaluate expectation values, which we
shall refer to as the one- and two-body bra-state equations,
\begin{eqnarray}
k\gamma_{-\vec{q}}(1-2s_{\vec{q}}\tilde{s}_{\vec{q}})+2\tilde{s}_{\vec{q}}- 2k\gamma_{-\vec{q}}\sum_{\vec{r}}\tilde{s}_{\vec{r}}^{(1)}s_{\vec{r}}^{(1)}-4\sum_{\vec{r}}e^{-i\vec{q}\cdot\vec{r}}s_{\vec{r}}\tilde{s}_{\vec{r}}^{(1)}&=&0,\\
k\tilde{s}_{1}\frac{1}{z}\sum_{\vec{\rho}}\delta_{\vec{r},\vec{\rho}} -2ks_{1}\tilde{s}_{\vec{r}}^{(1)} & = & 0\implies\nonumber \\
\tilde{s}_{\vec{r}}^{(1)} & = & \frac{\tilde{s}_{1}}{2zs_{1}}\delta_{|\vec{r}|,1}.
\end{eqnarray}

The remaining coefficients can be solved easily; we find
\begin{equation}
s_{\vec{q}}=\frac{1}{k\gamma_{-\vec{q}}}\left(1-\sqrt{1+k^{2}|\gamma_{\vec{q}}|^{2}\left[1-s_{1}/k\right]}\right),
\end{equation}
together with a self-consistency condition for $s_{1}$,
\begin{eqnarray}
s_{1} & = & \frac{1}{|\mathcal{A}|}\int_{\mathcal{A}}d\vec{q}\,\gamma_{-\vec{q}}s_{\vec{q}}\nonumber \\
 & = & \frac{1}{k|\mathcal{A}|}\int_{\mathcal{A}}d\vec{q}\,\left(1-\sqrt{1+k^{2}|\gamma_{\vec{q}}|^{2}\left[1-s_{1}/k\right]}\right)\,.
\end{eqnarray}
For the bra-state coefficients we have 
\begin{eqnarray}
\tilde{s}_{\vec{q}} & = & -\frac{\gamma_{-\vec{q}}}{2\sqrt{1+k^{2}|\gamma_{\vec{q}}|^{2}\left[1-s_{1}/k\right]}}\left(k-\tilde{s}_{1}\right),
\end{eqnarray}
where the value of $\tilde{s}_{1}$ can be evaluated directly,

\begin{align}
\tilde{s}_{1} & =\frac{1}{|\mathcal{A}|}\int_{\mathcal{A}}d\vec{q}\,\gamma_{\vec{q}}\tilde{s}_{\vec{q}}=-\frac{kI_{1}}{1-I_{1}},\\
I_{1} & =\frac{1}{|\mathcal{A}|}\int_{\mathcal{A}}d\vec{q}\frac{|\gamma_{\vec{q}}|^{2}}{2\sqrt{1+k^{2}|\gamma_{\vec{q}}|^{2}\left[1-s_{1}/k\right]}}.
\end{align}
The order parameter for the problem is the ``sub-lattice magnetisation'',
the average $z$-component of the magnetisation in one of the sub-lattices
(the total magnetisation is zero), $M_z=\frac{2}{N} \sum_{\vec i}S^z_{\vec i}$.
We use the fact that the expectation value of an operator $O$ in the CCM
approximation is given by
\begin{equation}
\langle O \rangle=\braOket{\tilde{\Psi}}{O}{\Psi}=
\braOket{\Phi_0}{(1+\tilde S)e^{-S}Oe^S}{\Phi_0}.
\end{equation}
In the SUB1 approximation we find the simple expression
\begin{equation}
\langle M_{z}\rangle_{B}=\frac{1}{2}\left(1-2\frac{1}{|\mathcal{A}|}\int_{\mathcal{A}}d\vec{q}\, s_{\vec{q}}\tilde{s}_{\vec{q}}+\frac{1}{k}\tilde{s}_{1}\right).
\end{equation}

\subsection{Super-SUB1 Approxmation\label{App:A-Super}}
Using the sublattice Fourier-transform, we can write the one-body equation
as (with $s_{\vec{\rho}}^{(1)}\equiv s_{1}^{(1)}$ using the lattice
symmetry) 
\begin{equation}
t\left(\gamma_{\vec{q}}-\,\gamma_{-\vec{q}}\, s_{\vec{q}}^{2}+s_{1}^{(1)}\gamma_{\vec{q}}\right)+\dfrac{U}{z}s_{\vec{q}}=0,\label{eq:ket}
\end{equation}
and the one-body bra-state equation is given by 
\begin{equation}
t\Bigl(1-2\langle\tilde{s}^{(1)}s^{(1)}\rangle\Bigr)\gamma_{-\vec{q}}-2t\,\gamma_{-\vec{q}}\,\tilde{s}_{\vec{q}}s_{\vec{q}}+\dfrac{U}{z}\Bigl(\tilde{s}_{\vec{q}}-2\left(s\ast\tilde{s}^{(1)}\right)_{\vec{q}}\Bigr)=0,\label{eq:bra}
\end{equation}
where, 
\begin{align}
\langle\tilde{s}^{(1)}s^{(1)}\rangle & =\sum\limits _{\vec{r}}\tilde{s}_{\vec{r}}^{(1)}s_{\vec{r}}^{(1)}=\dfrac{1}{|\mathcal{A}|}\int\limits _{\mathcal{A}}\tilde{s}_{\vec{q}}^{(1)}s_{\vec{q}}^{(1)}\, d\vec{q},\\
\left(s\ast\tilde{s}^{(1)}\right)_{\vec{q}} & =\sum\limits _{\vec{r}}\tilde{s}_{\vec{r}}^{(1)}s_{\vec{r}}\, e^{-\text{i}\vec{r}\cdot\vec{q}}=\dfrac{1}{|\mathcal{A}|^2}\int\limits _{\mathcal{A}}\int\limits _{\mathcal{A}}\tilde{s}_{\vec{q}_{2}}^{(1)}s_{\vec{q}_{1}}\delta_{\vec{q}_{1}-\vec{q}_{2},\vec{q}}^{\text{latt}}\, d\vec{q}_{1}d\vec{q}_{2}.
\end{align}
Here all $\vec{q}$'s are vectors defined within this first Brillouin
zone. The lattice delta function $\delta^{\text{latt }}$ defines
equality when both its arguments are translated back into the first
Brillouin zone. 

Solving both ket and bra equations (\ref{eq:ket},\ref{eq:bra}) results
in 
\begin{align}
s_{\vec{q}}=\dfrac{1}{k\gamma_{-\vec{q}}}\Bigl(1-\sqrt{1+k^{2}\,(1+s_{1}^{(1)})\,|\gamma_{\vec{q}}|^{2}}\Bigr),\label{eq:sq}
\end{align}
and 
\begin{align}
\tilde{s}_{\vec{q}} & =\dfrac{4\left(s\ast\tilde{s}^{(1)}\right)_{\vec{q}}-k\,\gamma_{-\vec{q}}\Bigl(1-2\langle\tilde{s}^{(1)}s^{(1)}\rangle\Bigr)}{2\Bigl(1-k\,\gamma_{-\vec{q}}\, s_{\vec{q}}\Bigr)}\nonumber \\
 & =\dfrac{4\left(s\ast\tilde{s}^{(1)}\right)_{\vec{q}}-k\,\gamma_{-\vec{q}}\Bigl(1-2\langle\tilde{s}^{(1)}s^{(1)}\rangle\Bigr)}{2\sqrt{1+k^{2}\,(1+s_{\vec{q}}^{(1)})\,|\gamma_{\vec{q}}|^{2}}}.
\end{align}
In the super-SUB1 approximation, we do not solve for $s_{\vec{q}}^{(1)}$
and $\tilde{s}_{\vec{q}}^{(1)}$, but replace them with the solution
to the unrestricted SUB2 solution of the $XXZ$ model, 
\begin{align}
\alpha_{\vec{q}}^{\Delta} & =\dfrac{K}{\gamma_{-\vec{q}}}\Bigl(1-\sqrt{1-\kappa^{2}\,|\gamma_{\vec{q}}|^{2}}\Bigr)\,,\\
\tilde{\alpha}_{\vec{q}}^{\Delta} & =\dfrac{D}{4K}\dfrac{\gamma_{-\vec{q}}}{\sqrt{1-\kappa^{2}\,|\gamma_{\vec{q}}|^{2}}}\,.
\end{align}
The coefficients $\kappa$, $K$ and $D$ all depend on $\alpha_{1}^{\Delta}$, \begin{align}
\kappa^{2} & =\dfrac{1+2\Delta\,\alpha_{1}^{\Delta}+2\left(\alpha_{1}^{\Delta}\right)^{2}}{(\Delta+2\Delta\,\alpha_{1}^{\Delta})^{2}},\\
K & =\Delta+2\alpha_{1}^{\Delta},\\
D^{-1} & =\dfrac{1}{|\mathcal{A}|}\int\limits _{\mathcal{A}}\dfrac{1-|\gamma_{\vec{q}}|^{2}/2}{\sqrt{1-\kappa^{2}\,|\gamma_{\vec{q}}|^{2}}}d\vec{q}-\dfrac{1}{2}.
\end{align}
We thus need to solve these equations self-consistently.

Having found $\alpha$, we approximate $\left(s\ast\tilde{s}^{(1)}\right)_{\vec{q}}$ by 
\begin{align}
\left(s\ast\tilde{\alpha}^{\Delta}\right)_{\vec{q}} & =\dfrac{1}{|\mathcal{A}|}\int\limits _{\mathcal{A}}\tilde{\alpha}_{\vec{q}'-\vec{q}}^{\Delta}\, s_{\vec{q}^{\prime}}\, d\vec{q}^{\prime}\\
 & =\dfrac{D}{4Kk}\dfrac{1}{|\mathcal{A}|}\int\limits _{\mathcal{A}}\dfrac{\gamma_{\vec{q}-\vec{q}^{\prime}}}{\gamma_{-\vec{q}^{\prime}}}\left(\dfrac{1-\sqrt{1+k^{2}\,\left(1+\alpha_{1}^{\Delta}\right)\,|\gamma_{\vec{q}^{\prime}}|^{2}}}{\sqrt{1-\kappa^{2}\,|\gamma_{\vec{q}-\vec{q}^{\prime}}|^{2}}}\right)\, d\vec{q}^{\prime}.
\end{align}
Since the full solution $\tilde{\alpha}^\Delta$ is a periodic function on the lattice, it
already incorporates the lattice delta function, and we can drop it
in the calculation. The sublattice magnetization of the Hubbard model
is, in this approximation, a function of the parameter $\Delta$,
and is given by \cite{bishop1991coupledcluster}
\begin{align}
\langle M_{\Delta}\rangle_{B} & =\frac{1}{2}-\sum\limits _{\vec{r}}\tilde{s}_{\vec{r}}\, s_{\vec{r}}-\sum\limits _{\vec{r}}\tilde{\alpha}_{\vec{r}}^{\Delta}\alpha_{\vec{r}}^{\Delta}.
\end{align}
Now, we can employ our knowledge about the staggered magnetization
of the $XXZ$ model, 
\begin{align}
M_{\Delta}^{\text{XXZ}}=\frac{1}{2}-\sum\limits _{\vec{r}}\tilde{\alpha}_{\vec{r}}^{\Delta}\alpha_{\vec{r}}^{\Delta},
\end{align}
to express the sub-lattice magnetization of Hubbard model as 
\begin{alignat}{2}
\langle M_{\Delta}\rangle_{B} & = & -\dfrac{D}{2Kk^{2}}\dfrac{1}{|\mathcal{A}|^{2}} & \int\limits _{\mathcal{A}}\int\limits _{\mathcal{A}}\dfrac{\gamma_{\vec{q}-\vec{q}^{\prime}}}{\gamma_{\vec{q}}\gamma_{-\vec{q}^{\prime}}}\left(\dfrac{1}{\sqrt{1+k^{2}\,(1+\alpha_{1}^{\Delta})\,|\gamma_{\vec{q}}|^{2}}}-1\right)\nonumber \\
 &  &  & \times\dfrac{1-\sqrt{1+k^{2}\,(1+\alpha_{1}^{\Delta})\,|\gamma_{\vec{q}^{\prime}}|^{2}}}{\sqrt{1-\kappa^{2}\,|\gamma_{\vec{q}-\vec{q}^{\prime}}|^{2}}}\, d\vec{q}\: d\vec{q}^{\prime}\nonumber \\
 &  & +M_{\Delta}^{\text{XXZ}} & \dfrac{1}{|\mathcal{A}|}\int\limits _{\mathcal{A}}\dfrac{1}{\sqrt{1+k^{2}\,(1+\alpha_{1}^{\Delta})\,|\gamma_{\vec{q}}|^{2}}}d\vec{q}.
\end{alignat}

\section{The super-SUB1 equation and excitation energies\label{app:B}}
As explained in Ref.\ \cite{macdonald1988,oles_comment_1990}
it is a subtle process to derive the Heisenberg limit of the Hubbard model.
To summarize their ideas succinctly, we disentangle the Hubbard-model Hamiltonian as [please note, the operators $T_m$ are not CCM operators, but are defined in Ref]
\begin{equation} 
H/t= T+\frac{U}{t} V,\quad T=T_0+T_1+T_{-1},
\end{equation}
where the label on $T_m$  denotes the number of potential quanta  added by
each operator,
\begin{equation}
[V,T_m]=mT_m.
\end{equation}
We then perform a unitary transformation removing the coupling terms 
$T_{\pm 1}$ from the Hamiltonian. 
This transformed Hamiltonian takes the form, to first order in $t/U$,
\begin{equation}
H=\frac{U}{t}V+ T_0+\frac{t}{U}[T_{-1},T_{1}].
\end{equation}
For half filling, the states satisfying $V|\phi\rangle=0$ are exactly
those that map on spin states, with one electron on each site. These
states are also annihilated by $T_0$ and $T_{-1}$, and in the space of these states
only the term $\frac{t}{U}T_{-1}T_{1}$ contributes,  which, as discussed in
Ref.\ \cite{macdonald1988} is in the spin-state
model space the Heisenberg model Hamiltonian parametrised
in terms of fermion operators.

So what is the importance of this? It means that if we wish to borrow the
$S^{(2)}$ operator from the Heisenberg model in the Hubbard model, we
should in principle first perform the inverse unitary transformation on
the operator $S^{(2)}$. 

Let us be a bit more specific, which may help us understand the
situation better. The $T$ operators take the form
\begin{eqnarray}
T_{0} & = & -\sum_{\langle\vec{i}\vec{j}\rangle}
\biggl(a_{\vec{i},\text{\ensuremath{\uparrow}}}b_{\vec{j}\downarrow}(n_{\vec{i}\downarrow}h_{\vec j \uparrow}+h_{\vec i\downarrow}n_{\vec j\uparrow})
-a_{\vec{i}\downarrow}b_{\vec{j}\uparrow}\left(n_{\vec i\uparrow}h_{\vec j\downarrow}+h_{\vec i\uparrow}n_{\vec j\downarrow}\right)\nonumber\\&&
+a_{\vec{i}\text{\ensuremath{\downarrow}}}^{\dagger}b_{\vec{j} \uparrow}^{\dagger}(h_{\vec i\uparrow}n_{\vec j\downarrow}+n_{\vec i\uparrow}h_{\vec j\downarrow})
-a_{\vec{i}\uparrow}^{\dagger}b_{\vec{j}\downarrow}^{\dagger}\left(n_{\vec j\uparrow}h_{\vec i\downarrow}+h_{\vec j\uparrow}n_{\vec i \downarrow}\right)
\biggr),\\
T_{1} & = & -\sum_{\langle\vec{i}\vec{j}\rangle}\left(a_{\vec{i} \text{\ensuremath{\uparrow}}}b_{\vec{j} \downarrow}n_{\vec i\downarrow}n_{\vec j\uparrow}-a_{\vec{i} \downarrow}b_{\vec{j} \uparrow}n_{\vec i\uparrow}n_{\vec j\downarrow}+a_{\vec{i} \text{\ensuremath{\downarrow}}}^{\dagger}b_{\vec{j} \uparrow}^{\dagger}h_{\vec i\uparrow}h_{\vec j \downarrow}-a_{\vec{i} \uparrow}^{\dagger}b_{\vec{j} \downarrow}^{\dagger}h_{\vec j\uparrow}h_{\vec i\downarrow}\right),\\
T_{-1} & = & -\sum_{\langle\vec{i}\vec{j}\rangle}\left(a_{\vec{i} \text{\ensuremath{\uparrow}}}b_{\vec{j} \downarrow}h_{\vec i \downarrow}h_{\vec j\uparrow}-a_{\vec{i} \downarrow}b_{\vec{j} \uparrow}h_{\vec i \uparrow}h_{\vec j \downarrow}+a_{\vec{i} \text{\ensuremath{\downarrow}}}^{\dagger}b_{\vec{j} \uparrow}^{\dagger}n_{\vec i \uparrow}n_{\vec j \downarrow}-a_{\vec{i} \uparrow}^{\dagger}b_{\vec{j} \downarrow}^{\dagger}n_{\vec j \uparrow}n_{\vec i\downarrow}\right),
\end{eqnarray}
where $h_\alpha=1-n_\alpha$, and $n_\alpha$ is the fermion number operator for
a given position and spin.
The unitary transformation on the Hubbard Hamiltonian takes the form
\begin{equation}
H_\text{equiv}=\mathcal{U} H \mathcal{U}^\dagger,
\end{equation}
with 
\begin{equation}
\mathcal{U}=\exp\left(\frac{t}{U}\left(T_{1}-T_{-1}\right)\right).
\end{equation}
To the dominant order in $t/U$, we find that the Hamiltonian takes the form
\begin{equation}
H_\text{equiv}=t\left( \frac{U}{t} V+T_0+\frac{t}{U}[T_{-1},T_{1}]\right).
\end{equation}
Clearly we will have to consider the subspace of smallest $V$ if $U$ gets large.
For half filling, this is the subspace annihilated by $V$, which is exactly the
subspace that maps onto spin states, i.e., which single fermion occupancy at
each site. States in this space are also annihilated by $T_0$ and $T_{-1}$, so that the only term remaining is the Heisenberg Hamiltonian
\begin{equation}
H_\text{he}=t\left( \frac{t}{U}T_{-1}T_{1}\right).\label{eq:HeiFerm}
\end{equation}

If we now apply the CCM method to Eq.\ (\ref{eq:HeiFerm}), we see that
the equations are different than the ones we get for the Hubbard
model--there is some similarity, but there are additional terms if we
compare them within the spin-state subspace. That can be most easily
understood in terms of inequivalent operators: the $S$ operators for
the fermion-version of the Heisenberg model are related to those of
the Hubbard model by an inverse unitary transformation
\begin{equation}
S_{\text{Hubbard}}=\mathcal{U}^\dagger S_{\text{Heisenberg}} \mathcal{U}.
\end{equation}
We would like to take over these coefficients from the Heisenberg to
the Hubbard model in a super-SUB1 approximation; if we do not want to
loose the Heisenberg model correspondence we have to include more than
the lowest order transformation of $S^{(2)}$. If we make the
approximation that the $S^{(2)}$ of the Hubbard model equals that of
the Heisenberg one, we miss the fact that we need the first order
corrections in the Hubbard version to reproduce the Heisenberg
results.  The idea is that we require that the CCM equations of the
Hubbard model go over into the equations for the Heisenberg model in
the limit $t/U\rightarrow0$. Any term that is absent in the Hubbard
model calculation gets added in by hand.  For the ground-state
calculations that corresponds exactly to the super-SUB1 approximation
employed in this paper; there is an effect on the equations for the
coefficients, but \emph{not} on the energy expressions.

The situation is slightly more
subtle for the excited state calculation. Rather than performing a
lengthy calculation, we extract the lowest corrections from the
Heisenberg model result for the magnon excitation energy,
\begin{equation}
\omega_{\vec{q}}/t=\frac{t}{U}2z \left[1+\alpha_1 (2-z |\gamma_{\vec q}|^2)
\right]
\end{equation}
As shown in the main text, we already get a term proportional
$1+2\alpha_1z$ from the main evaluation; we just need to add the
missing term in perturbatively.  We need to be careful since it acts
on $\vec Q$, not on the relative momentum, but comparing to the Heisenberg
CCM equations shows that the right answer is the equation
\begin{align}
-t\,\sum_{\langle\vec{i},\vec{j}\rangle}^{N/2}\Bigl(\chi_{\vec{i,}\vec{i}_{1}}^{s}\,
s_{\vec{i}_{2},\vec{j}}+\chi_{\vec{i}_{2},\vec{i}}^{s}\,
s_{\vec{i}_{1},\vec{j}}\Bigr)+U\,\chi_{\vec{i}_{1},\vec{i}_{2}}^{s}
\Bigl(1-\delta_{\vec{i}_{1},\vec{i}_{2}}
\left[1+\frac{t^2}{U^2} \sum_{\vec r,\vec \rho} a_{\vec r}^\Delta\delta_{\vec i_1,\vec r-\vec\rho}\right]
\Bigr)=\omega^{s}\,\chi_{\vec{i}_{1},\vec{i}_{2}}^{s}.\label{eq:spingap5a_app}
\end{align}
This of course means that this equation is no longer valid for small
$U/t$--but that should not come as a surprise since the super-SUB1
approximation also fails for such parameters.

\bibliographystyle{apsrev4-1}
\bibliography{paper1}

\end{document}